\documentclass[graybox,nospthms]{svmult}

\usepackage{mathptmx}       
\usepackage{helvet}         
\usepackage[scaled]{inconsolata}        
\usepackage{type1cm}        
\usepackage{makeidx}         
\usepackage{graphicx}        
\usepackage{multicol}        
\usepackage[bottom]{footmisc}

\usepackage{alltt}      
\usepackage{amsfonts}   
\usepackage{amsmath}    

\usepackage{amssymb}

\usepackage{amsthm}     
\usepackage{bussproofs}
\EnableBpAbbreviations
\usepackage{cite}
\usepackage{combelow}
\usepackage{cprotect}
\usepackage{datetime}
\usepackage[inline]{enumitem}
\usepackage[colorlinks,linkcolor=black,citecolor=black,urlcolor=black]{hyperref}
\usepackage[T1]{fontenc}
\usepackage[utf8]{inputenc}
\usepackage{makeidx}
\usepackage{mathpartir}
\usepackage{multirow}
\usepackage{newunicodechar}
\usepackage{proof}
\usepackage{stmaryrd}
\usepackage{tikz}
\usepackage{tikz-cd}
\usepackage{tikz-qtree}
\usepackage{upgreek}
\usepackage{varwidth}
\usepackage{xspace}
\usepackage[all]{xy}
\usepackage[all]{xypic}

\makeindex

\newunicodechar{∷}{::}
\newunicodechar{∀}{\ensuremath{\forall}}
\newunicodechar{∃}{\ensuremath{\exists}}
\newunicodechar{Σ}{\ensuremath{\Sigma}}
\newunicodechar{Γ}{\ensuremath{\Gamma}}
\newunicodechar{Δ}{\ensuremath{\Delta}}
\newunicodechar{∨}{\ensuremath{\lor}}
\newunicodechar{∧}{\ensuremath{\land}}
\newunicodechar{¬}{\ensuremath{\neg}}
\newunicodechar{λ}{\ensuremath{\lambda}}
\newunicodechar{∈}{\ensuremath{\in}}
\newunicodechar{∉}{\ensuremath{\not\in}}
\newunicodechar{∗}{\ensuremath{\ast}}
\newunicodechar{ℕ}{\ensuremath{\mathbb{N}}}
\newunicodechar{ℤ}{\ensuremath{\mathbb{Z}}}
\newunicodechar{ℚ}{\ensuremath{\mathbb{Q}}}
\newunicodechar{ℝ}{\ensuremath{\mathbb{R}}}
\newunicodechar{ℂ}{\ensuremath{\mathbb{C}}}
\newunicodechar{≤}{\ensuremath{\le}}
\newunicodechar{≺}{\ensuremath{\prec}}
\newunicodechar{∧}{\ensuremath{\wedge}}
\newunicodechar{⇒}{\ensuremath{\Rightarrow}}
\newunicodechar{→}{\ensuremath{\to}}
\newunicodechar{≠}{\ensuremath{\neq}}
\newunicodechar{⋯}{\ensuremath{\cdots}}
\newunicodechar{α}{\ensuremath{\alpha}}
\newunicodechar{β}{\ensuremath{\beta}}
\newunicodechar{γ}{\ensuremath{\gamma}}
\newunicodechar{δ}{\ensuremath{\delta}}
\newunicodechar{ε}{\ensuremath{\varepsilon}}
\newunicodechar{φ}{\ensuremath{\phi}}
\newunicodechar{×}{\ensuremath{\times}}
\newunicodechar{₀}{\ensuremath{_{0}}}
\newunicodechar{₁}{\ensuremath{_{1}}}
\newunicodechar{₂}{\ensuremath{_{2}}}
\newunicodechar{₃}{\ensuremath{_{3}}}
\newunicodechar{₄}{\ensuremath{_{4}}}
\newunicodechar{₅}{\ensuremath{_{5}}}
\newunicodechar{₆}{\ensuremath{_{6}}}
\newunicodechar{₇}{\ensuremath{_{7}}}
\newunicodechar{₈}{\ensuremath{_{8}}}
\newunicodechar{₉}{\ensuremath{_{9}}}
\newunicodechar{⟨}{\ensuremath{\langle}}
\newunicodechar{⟩}{\ensuremath{\rangle}}
\newunicodechar{⟦}{\ensuremath{\llbracket}}
\newunicodechar{⟧}{\ensuremath{\rrbracket}}

\DeclareSymbolFont{letters}{OML}{txmi}{m}{it}

\usetikzlibrary{petri} 
\usetikzlibrary{er}    
\usetikzlibrary{positioning,calc}

\makeatletter
\providecommand*{\toclevel@titlech}{0}
\edef\toclevel@authorch{\the\numexpr\toclevel@titlech+3}
\makeatother

\theoremstyle{plain}

\theoremstyle{definition}

\newcommand\authorsandsecondreaders[2]{\authorrunning{#1}\tocauthor{#1}\author{#1 \\~\\~\\ Second readers: #2}}

\DeclareSymbolFont{letters}{OML}{txmi}{m}{it} 
\let\oldPhi=\Phi
\renewcommand\Phi{\mathrm{\oldPhi}}

\DeclareUnicodeCharacter{3C3}{\ensuremath{\sigma}}
\DeclareUnicodeCharacter{220F}{\ensuremath{\Pi}}
\DeclareUnicodeCharacter{03A0}{\ensuremath{\Pi}}
\DeclareUnicodeCharacter{3B5}{\ensuremath{\varepsilon}}
\DeclareUnicodeCharacter{2211}{\ensuremath{\Sigma}}
\DeclareUnicodeCharacter{2124}{\ensuremath{\mathbb{Z}}}
\DeclareUnicodeCharacter{2208}{\ensuremath{\in}}
\DeclareUnicodeCharacter{27F9}{\ensuremath{\Longrightarrow}}
\DeclareUnicodeCharacter{21D2}{\ensuremath{\Rightarrow}}
\DeclareUnicodeCharacter{03BB}{\ensuremath{\lambda}}

\begin{document}

\setcounter{chapter}{18}
\title{Formalization of Security}
\authorsandsecondreaders{Gilles Barthe}{Lennart Beringer and Andreas Lochbihler}
\authorrunning{Gilles Barthe}

\institute{%
Gilles Barthe \at Max Planck Institute for Security and Privacy, Bochum, Germany, and IMDEA Software Institute,
  Madrid, Spain, \email{gilles.barthe@mpi-sp.org,gilles.barthe@imdea.org}
}
%
%
\maketitle
\label{chap:formalization-of-security}

\newcommand{\gb}[1]{\textcolor{red}{GB: #1}}

\index{Barthe, Gilles}
\index{Beringer, Lennart}
\index{Lochbihler, Andreas}

\abstract{
Proof assistants are often used to validate that designs and implementations
meet their expected security properties. A further motivation for using proof
assistants is to support certification. This chapter focuses on their applications
to system security, language-based security, secure compilation, and cryptography.
}
\bigskip
\bigskip

\section{Introduction}

\index{security|(}

Security is often a major consideration in the design and
implementation of software systems. Because security goals can be
difficult to achieve and security analyses can contain subtle flaws,
proof assistants are often used to validate that designs and
implementations meet their expected security properties. A further
motivation for using proof assistants is to support certification\index{certification},
e.g., Common Criteria\index{Common Criteria} evaluations, with mechanized proofs.

In this chapter, we focus on applications of proof assistants to
system security, language-based security, secure compilation, and
cryptography. We briefly discuss other applications at the end of
the chapter.

\section{Information Flow}

\index{information flow|(}

A fundamental security goal is to prevent illegal information flows,
including leaking secrets on public channels (confidentiality\index{confidentiality|(}) or
corrupting high-integrity data with untrusted values (integrity\index{integrity|(}). This
entails analyzing how information is flowing during the lifetime of a
system or the execution of a program. In general, information flow
security considers lattices of security levels (for confidentiality
and for integrity) and controls the flow of information between
distinct security levels. The simplest examples of lattices are:
\begin{itemize}
\item Confidentiality: $(\{ \textit{Public}, \textit{Secret} \}{,}\; \textit{Public} \leq \textit{Secret})$;
\item Integrity: $(\{ \textit{Trusted}, \textit{Untrusted} \}{,}\; \textit{Trusted} \leq \textit{Untrusted})$.
\end{itemize}
Information flow policies are typically used to (dis)allow information
flows: for example, a baseline confidentiality policy is that secrets
do not flow to public components, whereas a baseline integrity policy
is that untrusted (i.e., potentially corrupted) components do not flow
to trusted components. These two policies are instances of
noninterference\index{noninterference}, an information flow policy introduced by Goguen and
Meseguer~\cite{DBLP:conf/sp/GoguenM82a}. In a follow-up work, Goguen
and Meseguer~\cite{DBLP:conf/sp/GoguenM84} provide a powerful
technique, called unwinding\index{unwinding}, that allows to reduce proofs of
noninterference to simpler lemmas about one-step execution. Broadly
speaking, unwinding lemmas are stated relative to an equivalence
relation $\sim$ on states, where two states are related by $\sim$ if
and only if they cannot be distinguished by an attacker. There are two
basic unwinding lemmas:
\begin{itemize}
\item The step-consistent lemma states that one-step execution of $\sim$-related states yields  $\sim$-related states.  Informally, if $s_1\leadsto s'_1$ and $s_2\leadsto s'_2$, then under some additional conditions, $s_1\sim s_2$ implies $s'_1\sim s'_2$.
\smallskip
\item The step-preserving lemma guarantees that one-step execution
  preserves $\sim$.  Informally, if $s\leadsto s'$, then under some
  additional conditions, $s\sim s'$.
\end{itemize}  
By combining the two lemmas, one can prove that execution traces with
$\sim$-related initial states yield $\sim$-related final states.

\index{confidentiality|)}
\index{integrity|)}

\subsection{Systems-Level Security}
Many systems-level security policies can be modeled as
noninterference\index{noninterference} policies, and have been a main target of
formalization.

Rushby~\cite{Rushby92noninterference} used the EHDM Verification
System, a predecessor of PVS, to mechanize unwinding\index{unwinding} lemmas in a more
complex setting of intransitive noninterference, where information is
allowed to flow through specific channels. Von \hbox{Oheimb}
\cite{DBLP:conf/esorics/Oheimb04} formalizes a variant of Rushby's
framework in Isabelle/HOL and uses the resulting formalization to
reason about the security of the Infineon SLE66 chip. More recently,
Bracevac et al.~\cite{DBLP:journals/afp/BracevacGGMST18} use
Isabelle/HOL to formalize MAKS (Modular Assembly Kit of Security
properties), a rich framework developed by Mantel to reason about
information flow policies~\cite{DBLP:phd/de/Mantel2004}.

There is also a substantial body of work that establishes information
flow properties of specific systems. Many works focus on operating
systems and virtualization platforms.  The seL4 project uses
Isabelle/HOL to show that seL4 enforces integrity, authority
confinement~\cite{DBLP:conf/itp/SewellWGMAK11}, and intransitive
noninterference~\cite{DBLP:conf/sp/MurrayMBGBSLGK13}.  Barthe et
al.~\cite{DBLP:conf/fm/BartheBCL11, DBLP:conf/ccs/BartheBCLP14} use
Rocq (formerly known as Coq) to prove memory isolation for a model of virtualization that
includes caches and a translation lookaside buffer. They isolate a
class of executions that do not leak information to an attacker
executing on another partition and with control over the cache
replacement policy and the scheduler.  Dam et
al.~\cite{DBLP:conf/ccs/DamGKNS13} use HOL4 to formally verify
information flow security for a simple separation kernel for
ARMv7. Their main result is stated as an equivalence between an ideal
model in which the security requirements hold by construction with a
real model that faithfully respects the system behavior. Li et
al.~\cite{DBLP:conf/sp/LiLGNH21} and Tao et
al.~\cite{DBLP:conf/sosp/TaoYLLNG21} use Rocq to prove confidentiality
and integrity guarantees for a model of the KVM hypervisor, both with
respect to a sequential and to a relaxed memory model.  Azevedo de
Amorim et al.~\cite{DBLP:journals/jcs/AmorimCDDHPPPT16} use Rocq to
model the SAFE architecture and prove that the SAFE machine enforces
noninterference. A follow-up work~\cite{DBLP:conf/sp/AmorimDGHPST15}
extends this approach to provide a generic framework to enforce a rich
set of micro-policies, and instantiates the approach to verify several
prominent examples of micro-policies. Nelson et
al.~\cite{DBLP:journals/sigops/NelsonBKTW20} provides a recent
overview of machine-checked proofs of noninterference for secure
systems.

Some works focus explicitly on web
browsers\index{web browsers}. Bohannon~\cite{Bohannon12} uses Rocq to define a core model
of the Firefox web browser and proves reactive
noninterference~\cite{DBLP:conf/ccs/BohannonPSWZ09}. Jang, Tatlock, and
Lerner~\cite{DBLP:conf/uss/JangTL12} use the principles of shim
verification to verify QUARK, a web browser structured similarly to
Google Chrome.

Kanav, Lammich, and Popescu~\cite{DBLP:conf/cav/KanavL014} and
Bauereiss, Pesenti Gritti, Popescu, and
Raimondi~\cite{DBLP:conf/sp/BauereissG0R17} use Isabelle for
developing a conference management system\index{conference management systems} and a social platform\index{social platforms} with
verified confidentiality guarantees. Both proofs make use of bounded
deducibility~\cite{DBLP:conf/cav/KanavL014}, a
framework that combines the benefits of proofs by unwinding\index{unwinding}, the
precision of nondeducibility, and support for declassification.

Some formalizations are specifically developed or strongly motivated
by security evaluations, such as Common Criteria\index{Common Criteria}. The Formavie\index{Formavie}
project~\cite{betarte2002formavie} developed a formal model of the
JavaCard virtual machine\index{JavaCard} in Rocq. The formalization was used in an
Common Criteria certification\index{certification} at the highest level:\ EAL7. Andronick,
Boutali, Ly, and
Paulin-Mohring~\cite{DBLP:conf/tphol/AndronickCL03,DBLP:conf/fm/AndronickCP05}
build on this model to reason about isolation properties.  Hardin,
Smith, and Young~\cite{DBLP:conf/acl2/HardinSY06} use ACL2 for modeling
the Rockwell Collins AAMP7G microprocessor in the context of a NSA
certification.

\subsection{Language-Based Security}

\index{language-based security|(}

Language-based security~\cite{DBLP:journals/jsac/SabelfeldM03} is an
approach to strengthen security of applications through programming
language tools. Language-based security is foundational: indeed, many
language-based mechanisms come with a proof that they enforce a
security property of interest. This makes language-based security a
natural target for mechanization. Often, mechanizations that target
language-based security leverage existing mechanizations of
programming languages, compilers, and program logics, described in
other chapters of the handbook.

\subsubsection*{Imperative features}
Hri\cb{t}cu et al.~\cite{DBLP:conf/sp/HritcuGKPM13} use Rocq to formalize a fine-grained dynamic enforcement mechanism to
enforce noninterference of programs with exceptions\index{exceptions}.

Silver et al.~\cite{DBLP:conf/ecoop/SilverHCHZ23} develop an extension
of interaction trees\index{interaction trees}~\cite{DBLP:journals/pacmpl/XiaZHHMPZ20} with
support for exceptions and build a Rocq library to reason about
information flow properties of programs and to prove soundness of
information flow type systems.

In a different vein, Azevedo de Amorim, Hri\cb{t}cu, and
Pierce~\cite{DBLP:conf/post/AmorimHP18} use Rocq to
formalize a core programming language with manual memory management,
and show that safe programs satisfy a noninterference property,
namely that programs do not modify or depend on unreachable parts of
the state.

\subsubsection{Java-Like Languages}
Some of the earliest mechanizations of language-based security focus on
sequential fragments of Java virtual machine\index{Java} bytecode\index{bytecode}. For instance, Barthe, Pichardie, and Rezk~\cite{DBLP:conf/esop/BarthePR07} use Rocq to formalize an information flow type system for a sequential fragment of the Java bytecode, and prove that typable programs are noninterfering. Wasserrab, Lohner, and Snelting~\cite{DBLP:conf/pldi/WasserrabLS09} use Isabelle/HOL to formalize information flow control for program dependence graphs and instantiate their framework to derive a proof of noninterference for a sequential fragment of Java bytecode.

\subsubsection{C-Like Languages} Amtoft et al.~\cite{DBLP:conf/post/AmtoftDZABHOC12} formalize an information flow analyzer for the SPARK\index{SPARK} language in Rocq. The formalization provides an operational semantics for a core fragment of the language, a relational program logic, and a proof of its soundness with respect to the operational semantics, and a checker that is proved sound with respect to the semantics and program logics. The formalization also generates correctness certificates for an external information flow analyzer. Constanzo, Shao, and Gu~\cite{DBLP:conf/pldi/CostanzoSG16} develop a separation logic\index{separation logic} for reasoning about information flow of C-like programs in Rocq.  Proofs in separation logic\index{separation logic} guarantee the existence of a suitable bisimulation between two runs of the program under verification, and thus noninterference.

\subsubsection{Higher-Order Languages} Nanevski, Banerjee, and Garg~\cite{DBLP:conf/sp/NanevskiBG11} introduce relational Hoare type theory\index{Hoare type theory}, a relational extension of Hoare type theory for reasoning about information-flow properties of stateful higher-order programs. Relational Hoare type theory uses a shallow embedding of programs into Rocq. Recently, Frumin, Krebbers, and Birkedal~\cite{DBLP:conf/sp/FruminKB21} use the Iris\index{Iris} framework to mechanize a concurrent separation logic\index{separation logic} to reason about timing-sensitive noninterference of concurrent higher-order stateful programs. In subsequent work, Gregersen, Bay, Timany, and Birkedal~\cite{DBLP:journals/pacmpl/GregersenBTB21} introduce modal weakest precondition and show how they can be used to reason modularly about the more permissive termination-sensitive noninterference.

\subsubsection{Concurrency and Probabilistic Choice}
Barthe and Prensa-Nieto~\cite{DBLP:journals/jcs/BartheN07} use
Isabelle/HOL to mechanize proofs of noninterference for a concurrent
imperative language.  Popescu, H\"olzl, and Nipkow use Isabelle/HOL to
mechanize proofs of noninterference for source languages with
concurrency\index{concurrency}~\cite{DBLP:conf/cpp/PopescuHN12} and probabilistic
choice~\cite{DBLP:conf/cpp/0001HN13}. Murray, Sison, Pierzchalski, and
Rizkallah~\cite{DBLP:conf/csfw/MurraySPR16} use Isabelle/HOL to prove
soundness for a value-dependent information flow analysis for a core
programming language with concurrency.  Motivated by proof automation,
Ernst and Murray~\cite{DBLP:conf/cav/ErnstM19} develop a concurrent
separation logic\index{separation logic} for reasoning about data-dependent information flow
policies of C-like programs. They implement their logic into an
automated prover, called SecC\index{SecC}, and use Isabelle\slash HOL to mechanize its
soundness proof.

\subsubsection{Dynamic Enforcement}
Most mechanizations consider static enforcement. However, there is a large body of the security literature that uses dynamic enforcement mechanisms. Beringer~\cite{DBLP:conf/aplas/Beringer12} formalizes a hybrid enforcement approach that combines static analysis and dynamic taint tracking. The approach is proved sound with respect to an operational semantics in Rocq. Recently, Xian and Chong~\cite{DBLP:conf/sp/XiangC21} develop a coarse-grained dynamic information flow system for a Java-like language, and mechanize the soundness of a core subsystem using Rocq. Also recently, Vassena et al.~\cite{DBLP:journals/pacmpl/VassenaRGRS19} use Agda  to establish an equivalence between two styles for dynamic enforcement of information-flow policies.


\subsubsection{Side Channels}

\label{side channels|(}

Information flow policies have also been used to reason about
side-channel leakage.  In this context, policies state that leakage is
independent of secrets and are defined on top of an instrumented
semantics that captures leakage. One popular policy is the so-called
(cryptographic) constant-time policy. It is based on a leakage model
where branching statements leak their guards, memory reads and writes
leak the addresses of the memory accessed, and (in some works)
variable-time arithmetic\index{arithmetic} instructions leak (size of) their
operands~\cite{DBLP:conf/sp/BarbosaBBBCLP21}. Building on top of the CompCert\index{CompCert}
formalization, Barthe et al.~\cite{DBLP:conf/ccs/BartheBCLP14} define
and formally verify a type system that enforces constant-time for
assembly programs. Cock~\cite{DBLP:journals/corr/abs-1211-6197} uses a
formalization of pGCL\index{pGCL} in Isabelle to obtain formally verified upper
bounds on programs leakage.

Resource\index{resources} usage (see below) is often a side channel that can be
exploited by an attacker to retrieve confidential information. Such
attacks can be avoided by ensuring that programs are
constant-resource---i.e., their resources consumptions do not depend
on secrets.  The constant-resource policy can be seen as an instance
of an observational information flow policy, and more specifically, a
noninterference policy with respect to a resource-instrumented
semantics of programs. Ngo, Dehesa-Azuara, Frederikson, and
Hoffmann~\cite{DBLP:conf/sp/NgoDFH17} define constant-resource type
systems and use Agda to prove soundness of the type systems with
respect to a resource-instrumented semantics.

\label{side channels|)}
\index{language-based security|)}
\index{information flow|)}

\section{Resource Usage}

\index{resources|(}

There is a large body of work that uses proof assistants for reasoning
about complexity bounds of popular algorithms and implementations. The
most direct approach for reasoning about the complexity of algorithms
is to prove formally an upper bound on a mathematical definition of
the cost function. This direct approach has been used for instance by
Nipkow~\cite{DBLP:conf/itp/Nipkow15} and by Eberl, Halsbeck, and
Nipkow~\cite{DBLP:journals/jar/EberlHN20} for proving upper bounds of
the cost of deterministic and probabilistic algorithms. However, it is
often better to formalize and instantiate generic tools, such as
master theorems. For instance, Eberl~\cite{DBLP:journals/jar/Eberl17}
formalizes the Akra–Bazzi\index{Akra--Bazzi method} method in Isabelle, and Tassarotti and
Harper formalize~\cite{DBLP:conf/itp/Tassarotti018} Karp's cookbook\index{Karp's theorem}
theorem for verifying tail bounds of randomized algorithms. Yet
another approach is to build general libraries for cost. For instance,
Danielsson~\cite{DBLP:conf/popl/Danielsson08} uses Agda to formalize a
library for reasoning about the time complexity\index{time complexity} of purely functional\index{functional programming}
data structures. Li, Xia, and
Weirich~\cite{DBLP:journals/pacmpl/LiXW21} define a shallow embedding
of the claivoyance monad\index{monads} in Rocq, and leverage this embedding to reason
formally about the cost of lazy evaluation. A related approach is to
build automated tools for reasoning about complexity of functions
with respect to a cost model. Early work by
Benzinger~\cite{DBLP:journals/jfp/Benzinger01} uses a combination of
abstract interpretation\index{abstract interpretation} and recurrence solvers to reason about the
cost of Nuprl expressions.

Similar techniques can also be used to reason about implementations.
Alternatively, one can formalize metatheoretical properties of static
analyses, type systems and program logics for cost, and use their
guarantees to reason about the cost of specific algorithms. For
instance, Cachera, Jensen, Pichardie, and
Schneider~\cite{DBLP:conf/fm/CacheraJPS05} implement a formally
verified algorithm for checking that JavaCard\index{JavaCard} programs do not allocate
memory in loops and therefore execute in bounded memory.  Aspinall et
al.~\cite{DBLP:journals/tcs/AspinallBHLM07} formalize the soundness of
a resource-aware program logic for a fragment of the Java virtual
machine\index{Java} using Isabelle. Carbonneaux, Hoffmann, and
Shao~\cite{DBLP:conf/pldi/Carbonneaux0S15} use Rocq to verify the
soundness of a program logic that supports compositional
potential-based resource analyses of programs. Later, Carbonneaux,
Hoffmann, Reps, and Shao~\cite{DBLP:conf/cav/Carbonneaux0RS17} develop
a resource bound analysis that generates Rocq certificates of their
correctness for a core language; one main advantage of their analysis
is that it supports polynomial bounds. Chargu\'eraud and
Pottier~\cite{DBLP:conf/itp/ChargueraudP15} formalize an approach
based on characteristic formulas and time credits to prove the
complexity of a union--find\index{union--find data structure} implementation. Later, Chargu\'eraud, and
Pottier~\cite{DBLP:journals/jar/ChargueraudP19} and M\'evel, Jourdan,
and Pottier~\cite{DBLP:conf/esop/GueneauCP18} formalize a separation
logic\index{separation logic} with time credits and negative time credits in Rocq and use their
logic to prove complexity of several algorithms, including the
union--find data structure. Pottier et
al.~\cite{DBLP:journals/pacmpl/PottierGJM24} extend these works to
support reasoning about thunks and credits and use the resulting
framework to prove complexity bounds for several functional data
structures.

In a probabilistic setting, H{\"{o}}lzl and
Nipkow~\cite{DBLP:journals/corr/abs-1212-3870} prove expected runtime
of the Zero\-Cond\index{ZeroCond} protocol in Isabelle. Later,
H{\"{o}}lzl~\cite{DBLP:conf/itp/Holzl16} formalize in Isabelle/HOL a
weakest pre-expectation calculus for expected running times based
on~\cite{DBLP:journals/jacm/KaminskiKMO18}. Tassarotti and
Harper~\cite{DBLP:journals/pacmpl/TassarottiH19} use the Iris\index{Iris}
framework in Rocq to formalize a concurrent separation logic\index{separation logic} for
probabilistic programs and use their framework to derive complexity
bounds for skiplists. Avanzini et
al.~\cite{DBLP:journals/pacml/AvanziniBGMV24} extend EasyCrypt
with an expectation logic for probabilistic programs
and use the logic to prove upper bounds for skiplists. We refer
to~\cite{DBLP:conf/atva/NipkowEH20} and~\cite[\S
  6.2]{DBLP:books/hal/Chargueraud23} for more detailed overviews of
formally verified complexity analysis.

\index{resources|)}

\section{Access Control and Capabilities}

\index{access control|(}

Access control is a classic mechanism to enforce security of computer
systems. Access control policies are typically defined by large sets
of rules. The interactions between different rules may have unintended
implications or simply lead to inconsistencies, when rules disagree
whether or not to grant access. Bad interactions between rules can
compromise security, making it important to analyze formally the
consequences and consistency of access control policies. Capretta,
Stepien, Felty, and Matwin~\cite{DBLP:conf/ccs/CaprettaSFM07} verify
using Rocq an algorithm for conflict detections in firewalls, and use
their algorithm to detect conflicts in policies with several hundred
thousands of rules. Following a similar approach, St-Martin and
Felty~\cite{DBLP:conf/cpp/St-MartinF16} verify formally in Rocq an
algorithm for detecting conflicts in eXtensible Access Control Markup
Language (XACML)\index{XACML} policies. Brucker, Br{\"{u}}gger, and
Wolff~\cite{DBLP:journals/stvr/BruckerBW15} use formalize in HOL
firewall policies, and the HOL-TESTGEN\index{HOL-TESTGEN} tool to generate abstract test
cases from the formal model. Sohr, Drouineaud, Ahn, and
Gogolla~\cite{DBLP:journals/tkde/SohrDAG08} verify in Isabelle/HOL
role-based access control (RBAC) policies. Their verification relies
on an Isabelle/HOL formalization of first-order linear temporal logic\index{linear temporal logic},
in which RBAC policies can be encoded. More recently, Cutler et
al.~\cite{cutler2024cedar} use Lean to formalize key properties of
Cedar\index{Cedar}, an expressive authorization language used by Amazon Web
Services\index{Amazon Web Services}. Cedar supports role-based, attribute-based, and
relation-based access control.

Capability-based systems support powerful mechanisms, such as
transferring capabilities, and are able to enforce fine grained
security policies in low-level systems. As such, they are an
interesting and challenging target for
mechanization. CHERI\index{CHERI}~\cite{cheri} uses unforgeable capabilities to
guarantee safety and security in presence of untrusted code. The CHERI
team
\cite{DBLP:conf/sp/NienhuisJBFR0NN20,DBLP:conf/esop/BauereissCSAESB22}
has produced several mechanizations of CHERI in Isabelle/HOL. These
formalizations establish monotonicity properties, ensuring for
examples that capabilities cannot be increased during execution.  In
addition, Park, Pai, and Melham~\cite{DBLP:conf/tacas/ParkPM23} and
Zaliva et al.~\cite{cheri-c-asplos} provide formal models of CHERI C
in Isabelle and Rocq respectively.

CERISE\index{CERISE}~\cite{DBLP:journals/pacmpl/GeorgesGSTTHDB21,DBLP:conf/csfw/StrydonckGGTTPB22} is a program logic for reasoning about capabilities in presence of
unknown code. The soundness of the CERISE program logic is established
in Rocq using logical relations.

\index{access control|)}

\section{Hyperproperties}

\index{hyperproperties|(}

Program verification is traditionally focused on trace properties,
including safety and liveness properties. However, many security
properties are hyperproperties~\cite{DBLP:journals/jcs/ClarksonS10},
i.e., sets of sets of traces, rather than properties. A special class
of hyperproperties is hypersafety, which guarantees that nothing goes
bad across a set of executions; for instance, noninterfence is a
typical example of hypersafety.
Hyperlogics\index{hyperlogics}~\cite{DBLP:journals/siglog/Finkbeiner23} are temporal or
program logics for verifying that programs satisfy a given
hyperproperty. In their most expressive forms, hyperlogics extend
usual logics by allowing for arbitrarily nested quantifications over
program traces.

Despite their importance in security, there are relatively few
formalizations that consider hyperproperties. Antonopoulos et
al.~\cite{DBLP:journals/pacmpl/AntonopoulosKLNNN23} introduce a
relational variant of Kleene Algebra with Tests, called BiKAT\index{BiKAT}, which
allows reasoning about for all there exists properties. They prove the
soundness of their approach using Rocq.  Dardinier
and M\"uller~\cite{DBLP:journals/pacml/pldi/DardinierM24} introduce
Hyper Hoare Logic\index{Hyper Hoare Logic}, and use Isabelle to prove
soundness and completeness of the proof system. Gladshtein et
al.~\cite{DBLP:journals/pacml/pldi/GladshteinZAAS24} introduce the
Logic for Graceful Tensor Manipulation\index{Graceful Tensor Manipulation},
a separation hyperlogic\index{separation logic} for
reasoning about structured data. Their logic is formalized in Rocq.

\index{hyperproperties|)}

\section{Secure Compilation}
\label{sec:secure-compilation}

\index{compilation|(}
\index{secure compilation|(}

Secure compilation is a broad area of research that aims to guarantee
that low-level programs output by compilers are secure. This entails
proving that low-level programs are protected against common forms of
vulnerabilities, which are typically modeled by safety policies, and
verify security policies, including information flow and resource
control. In principle, compilers can guarantee these protections by
mitigation or by preservation. In the first case, the compiler carries
analyses and passes that ensure the desired property, whereas in the
second case, the compiler assumes that the source program satisfies
the desired property, and ensures that the property is preserved by
compilation. Unfortunately, compilers are typically not designed with
security in mind, and there are documented examples of compilers that
do not enforce the stated policies and do not carry security
properties from source to generated programs, see for
instance~\cite{DBLP:conf/sp/DSilvaPS15,DBLP:conf/eurosp/SimonCA18}.
These examples illustrate the challenges of preserving security during
compilation.

\subsection{Definitional Works and General Frameworks}

Defining secure compilation is nontrivial: indeed, the classic notion
of refinement used in the definition of compiler correctness does not
preserve security properties. Abadi~\cite{DBLP:conf/icalp/Abadi98} was
among the first to popularize the idea that secure compilation could
be understood through the lens of full abstraction, a notion that was
originally introduced by Plotkin~\cite{DBLP:journals/tcs/Plotkin77}
for relating observational equivalence and denotational equality, and
used by Mitchell~\cite{DBLP:journals/scp/Mitchell93} to compare
notions of program equivalence. Specifically, given a notion of
equivalence on a source language and a notion of program equivalence
on a target language, a mapping from source to target programs is
fully abstract if and only if it preserves and respects
equivalence. Abadi~\cite{DBLP:conf/icalp/Abadi98} showcases the
potential and obstacles of full abstraction, and suggests other
alternative approaches, including preservation of security-type
systems.

Although full abstraction remains a desirable property for compilers,
Parrow \cite{DBLP:journals/mscs/Parrow16}, Gorla and Nestmann
\cite{DBLP:journals/mscs/GorlaN16}, and Patri\-gani and
Garg~\cite{DBLP:conf/csfw/PatrignaniG17} highlight some shortcomings
of this notion. In particular, Patr\-igani and
Garg~\cite{DBLP:conf/csfw/PatrignaniG17} show an example of a fully
abstract compiler that fails to preserve confidentiality.  In
addition, Patri\-gani and Garg~\cite{DBLP:conf/csfw/PatrignaniG17} put
forward a general criterion, called trace-preserving compilation, for
preserving all safety hyperproperties\index{hyperproperties}.  The landscape of secure
compilation is further explored by Abate et
al.~\cite{DBLP:conf/csfw/AbateB0HPT19}, and by Abate et
al.~\cite{DBLP:journals/toplas/AbateBCDGHPTT21}. These works introduce
different criteria for secure compilation and compare their relative
strengths. In many cases, they also provide sufficient conditions and
alternative characterizations for these criteria. Their general
framework and many of the notions have been formalized in Rocq and used
to reason about specific compilers.

Proof-carrying code\index{proof-carrying code}~\cite{DBLP:conf/popl/Necula97} popularized the
idea of generating machine-checked proofs that low-level programs
satisfy safety and security policies. Foundational proof-carrying
code~\cite{DBLP:conf/popl/AppelF00} is a form of proof-carrying code
that uses a proof assistant to minimize the trusted computing base
and prove that programs satisfy the required policies relative
to a machine-checked formalization of program semantics. They
illustrate their approach for a small language using the Twelf prover.
Syntactic foundational proof-carrying
code~\cite{DBLP:conf/lics/HamidSTMN02} is an alternative approach that
does not require building complex program semantics. In general, these
approaches are focused on trace properties rather than security
properties.

The aforementioned works consider programs in isolation. Abate et
al.~\cite{DBLP:conf/ccs/AbateABEFHLPST18} introduce secure
compartmentalizing compilation\index{compartmentalizing compilation}, a relative notion which ensures that
compilation does not increase the insecurity of programs with
compromised components. Compartmentalizing compilation notably differs
from other works of correct and secure compilation, whose guarantees
are traditionally restricted to safe source programs. They use Rocq to
verify secure compartmentalizing compilation from a core language with
unsafe behaviors to a simple machine with built-in
compartmentalization. Their work introduces new techniques, including
a recomposition lemma, that allows to structure compiler correctness
proofs. Later, El-Korashy et
al.~\cite{DBLP:conf/csfw/El-KorashyBTDGH22} establish a similar
result for a richer language for a language with mechanisms to share
and dereference safe pointers\index{pointers} across components. Their work also
introduces new simulation techniques that are used to extend the
recomposition lemma to this richer setting.

\subsection{Compiler-Based Enforcement and Mitigations}

The aforementioned efforts are targeted at understanding the complex
landscape of secure compilation. As such, their main results are not
tied to a specific compiler, nor to a specific security property. In
contrast, other works formally establish mitigation or enforcement for
specific settings.

Jinja\index{Jinja} \cite{DBLP:journals/tcs/KleinN03} and CertiCartes\index{CertiCartes} \cite{DBLP:conf/fase/BartheD04} are Isabelle and Rocq formalizations of
sequential fragments of the Java virtual machine\index{Java} and include
machine-checked proofs of correctness of the bytecode\index{bytecode} verifiers.

SoftBound\index{SoftBound}~\cite{DBLP:conf/pldi/NagarakatteZMZ09} is a LLVM-based\index{LLVM}
transform that enforces spatial safety of C programs by inserting
runtime bound checks. Its design is formalized in Rocq and validated
against an operational semantics of C programs.

ARMor\index{ARMor}~\cite{DBLP:conf/emsoft/ZhaoLSR11} is a certifying compiler that
ensures memory safety and control integrity of ARM code. ARMor uses a
formally verified program logic atop a semantics of ARM instructions
to discharge proof obligations within HOL.
RockSalt\index{RockSalt}~\cite{DBLP:conf/pldi/MorrisettTTTG12} is a formally verified
checker to enforce that binary programs respect Google's Native Client
(NaCl) policy. RockSalt is written in Rocq, and is
formally verified against an operational semantics of x86.
KCofi\index{KCofi}~\cite{DBLP:conf/sp/CriswellDA14} is a system that ensures
control flow integrity for commodity operating systems. Its design is
formalized in Rocq and proved correct relative to
an operational semantics of the KCofi virtual machine.
CompCertSFI\index{CompCertSFI}~\cite{DBLP:conf/esop/BessonBDJW19} is an extension of
CompCert\index{CompCert} that enforces correct sandboxing; it comes with a formal
proof that compiled programs verify the sandboxing policy.
CertrBPF\index{CertrBPF}~\cite{DBLP:conf/cav/YuanBTHZB22} uses the CompCert\index{CompCert} compiler
to generate a formally verified verifier for eBPF\index{eBPF} virtual machine.

Some security-enhancing transformations have complex security proofs
that go beyond the typical scope of secure compilation, by requiring,
e.g., probabilistic arguments. In this case, it is common to prove only
correctness of the transformation. For instance, Fournet, Keller, and
Laporte~\cite{DBLP:conf/csfw/FournetKL16} prove correctness of a
variant of CompCert\index{CompCert} with a backend to circuits used for verifiable
computation.  Monniaux~\cite{DBLP:conf/cpp/Monniaux24} extends the
CompCert compiler with support for stack canaries and pointer
authentication. He then uses simulations to prove that both passes
preserve the semantics of programs.

\subsection{Information Flow and Resource Control Policies}

One specific line of research within secure compilation focuses on
preservation of side-channel countermeasures. It is well known that
mainstream compilers break constant-time by introducing branching
statements, or to a lesser extent secret-dependent memory accesses. On
the other hand, preservation of side-channel countermeasures would
allow developers to use existing analysis and mitigation tools, which
are often developed for source programs or intermediate
representations, without worrying about potential issues introduced by
the compiler. A perhaps surprising outcome is that existing verified
compilers mainly preserve constant-time.  In particular, there are
mechanized proofs that a slightly modified version of the CompCert\index{CompCert}
compiler and the Jasmin\index{Jasmin} compiler (Section~\ref{sssec:jasmin}) preserve
constant-time~\cite{DBLP:conf/csfw/BartheGL18,DBLP:journals/pacmpl/BartheBGHLPT20,DBLP:conf/ccs/BartheGLP21}. Recently,
Arranz-Olmos et al.~\cite{DBLP:journals/tches/OlmosBGGLLOS24} extend
the Jasmin compiler with stack zeroization, a security of
countermeasure that overwrites data before returning from a sensitive
computation. Stack zeroization guarantees absence of leakage in a
stronger attacker model in which an attacker sees the values of the
stack upon function return.

Another specific line of research within secure compilation focuses on
preservation of resource consumption. The goal here is mainly to show
that resource usage can be estimated (upper-bounded) correctly at
source level. An alternative is to estimate directly the cost of
generated code without analyzing the source code. Blazy, Maronese, and
Pichardie~\cite{DBLP:conf/vstte/BlazyMP13} formalize a worst-case
execution time (WCET) analysis on top of the CompCert\index{CompCert} compiler.
Amadio et al.~\cite{DBLP:conf/fopara/AmadioABBCGMMMPPRCST13} develop
an end-to-end framework to prove space and time bounds for an 8-bit
CPU. The core of their framework is a formally verified compiler,
which guarantees that the results of the source analysis are sound
with respect to a cost model of machine code\index{machine code}. Following a similar
approach, Carbonneaux, Hoffmann, Ramananandro, and
Shao~\cite{DBLP:conf/pldi/Carbonneaux0RS14} instrument the CompCert\index{CompCert}
compiler to prove stack-space bounds for machine code. Their
formalization includes a quantitative program logic to reason about
stack-space bounds at source level and a certified transformer that
turns the bounds obtained by source-level reasoning into valid bounds
for machine code\index{machine code}.  Besson, Blazy, and
Wilke~\cite{DBLP:conf/itp/BessonBW17} prove a similar result using a
more precise memory model for CompCert\index{CompCert}; Wang, Wilke, and
Shao~\cite{DBLP:journals/pacmpl/WangWS19} further refine their
approach to support fine-grained stack policies. Similar results have
been formally verified for other certified compilers including
CakeML\index{CakeML}~\cite{DBLP:journals/pacmpl/Gomez-LondonoPS20} and
Jasmin\index{Jasmin}~\cite{DBLP:conf/ccs/AlmeidaBBBGLOPS17} compilers. More
recently, Paraskevopoulou and
Appel~\cite{DBLP:journals/pacmpl/Paraskevopoulou19} use Rocq to prove
that closure conversion is safe for space (and time) bounds. One key
novelty of their work is a notion of logical relation that is
compatible with garbage collection.

Finally, several works consider secure compilation to capability-based
machines.  For instance, Georges, Trieu, and
Birkedal~\cite{DBLP:journals/pacmpl/GeorgesTB22} use the Iris\index{Iris}
framework for proving full abstraction for the ``overlay'' semantics
of a capability-based language.

\index{compilation|)}
\index{secure compilation|)}

\section{Cryptography}
\label{sec:cryptography}

\index{cryptography|(}

Cryptography is an essential component for building secure systems, and
arguably one that comes with the strongest mathematical guarantees. It is
therefore highly desirable to formally verify claims of security guarantees of
cryptographic designs.

\subsection{Security Proofs in Computational Model}

Goldwasser and Micali~\cite{DBLP:journals/jcss/GoldwasserM84} introduce the
computational model, which underlies the overwhelming majority of modern
cryptographic proofs. In this model, adversaries are probabilistic computations
with oracle accesses and their interactions with cryptographic systems are also
described by probabilistic computations called security experiments. Such
experiments are used to measure the security of a cryptographic system;
typically, one wants to show that every ``resource-bounded'' adversary has a
``small'' probability of winning the security experiment. Making this statement
precise requires one to define resource-bounded adversaries, and leads to
different settings. Broadly speaking, cryptographic proofs are typically
carried in one of two settings:\ information-theoretic or computational.

In the information-theoretic setting, one restricts the number of oracle
queries that can be performed by the adversary, and one upper-bounds the
winning probability of an adversary by an algebraic expression that depends on
the number of oracle queries. In the computational setting, one additionally
restricts the computational power of the adversary, and one reduces security of
the cryptographic construction to the security of a problem that is assumed to
be computationally hard. These reductionist statements are typically of the
form: for every adversary $\mathcal{A}$ against the security of the
cryptographic scheme $\mathbf{S}$, there exists a solver $\mathcal{B}$ for some
hard problem $\mathbf{P}$ such that the winning probability $p_\mathcal{A}$ of
$\mathcal{A}$ is upper-bounded as a function of the winning probability
$p_\mathcal{B}$ of $\mathcal{B}$. Moreover, the execution time $t_\mathcal{B}$
of $\mathcal{B}$ is upper-bounded by a function of the execution time
$t_\mathcal{A}$ of $\mathcal{A}$. In the ideal setting, $p_\mathcal{A}\leq
p_\mathcal{B}+\epsilon$, and $t_\mathcal{B}\leq t_\mathcal{A}+\delta$, with
$\epsilon$ and $\delta$ being very small. In this case, the reduction is tight.

However, there are sometimes multiplicative factors that make the reduction
looser. Note that in general, $\mathcal{B}$ invokes $\mathcal{A}$ as a
subroutine, and both $\delta$ and $\epsilon$ will depend on the number of
oracle queries that can be performed by the adversaries. Further note that
there may be several reductions, which for instance make different complexity
trade-offs---smaller $\epsilon$, larger $\delta$. Moreover, the reductions can
involve multiple hard problems. Finally, note that our exemplary reductionist
statement above expresses concrete bounds, which may be more relevant for
practical purposes. However, cryptographers also like to reason about
asymptotic security, in which cases one requires that $\mathcal{A}$ and
$\mathcal{B}$ execute in probabilistic polynomial-time, and prove that the
winning probability of $\mathcal{A}$ is negligible, assuming that the winning
probability of $\mathcal{B}$ is negligible. In all cases, the notions of
polynomial-time and negligibility are set relative to a security parameter by
which the experiments are (uniformly) parameterized.

Reductionist proofs are complex and error-prone. To tame their the complexity,
cryptographers~\cite{DBLP:conf/eurocrypt/BellareR06,DBLP:journals/iacr/Shoup04}
have developed and adopted a code-based approach, where probabilistic
experiments are written as probabilistic programs, and reductionist proofs are
decomposed into a sequence of small steps. In an inspirational work,
Halevi~\cite{DBLP:journals/iacr/Halevi05a} suggests the possibility of
mechanizing these proofs.

\subsubsection{CertiCrypt and EasyCrypt}

\index{CertiCrypt|(}
\index{EasyCrypt|(}

CertiCrypt~\cite{DBLP:conf/popl/BartheGB09} is among the first and most
complete formalizations of provable security in a proof assistant---see also
Affeldt et al.~\cite{DBLP:conf/provsec/AffeldtTM07} and Nowak
\cite{DBLP:conf/icics/Nowak07} for contemporary but less developed efforts. In
a nutshell, CertiCrypt is a Rocq library that formalizes many ideas and
techniques of the computational model. First of all, CertiCrypt provides a deep
embedding of a probabilistic language with adversarial computations, with a
cost-instrumented semantics for modeling complexity. Then, CertiCrypt provides
a rich set of tools for reasoning about adversarial computations. The main tool
is an expressive program logic pRHL\index{pRHL} (probabilistic relational Hoare logic) for
relating two programs with respect to relational pre-~and postconditions, both
modeled as shallow binary relations on program memories. Another tool is a
nonrelational program logic for upper-bounding the probability of events in
output distributions. Other tools include proof principles and program
transformations based on dependence and dataflow analysis\index{dataflow analyses}, and specific tactics
for cryptography, including a tactic for interprocedural code motion of
probabilistic assignments (also known as eager and lazy sampling) and
conditional equivalence (also known as equivalence up to failure event).
CertiCrypt has been used for verifying several classic examples from provable
security, including encryption schemes, signature schemes, hash functions, and
zero-knowledge\index{zero-knowledge proofs} proofs.

One advantage of CertiCrypt is that it is developed as a Rocq library and
therefore offers direct access to existing Rocq developments. For instance,
Barthe et al.~\cite{DBLP:journals/jcs/BartheGHOB13} build on Th\'ery and
Hanrot's formalization of elliptic curves\index{elliptic curves}~\cite{DBLP:conf/tphol/TheryH07} to
prove indifferentiability of a hash function into elliptic curves. Similarly,
Almeida et al.\ \cite{DBLP:conf/ccs/AlmeidaBBD13} uses the
CompCert\index{CompCert} compiler~\cite{DBLP:journals/cacm/Leroy09} to carry
the security proof of RSA-OAEP to an assembly-level implementation. In
addition, this work develops a technique to check that CompCert\index{CompCert}
does not create timing side channels by introducing branching on secrets during
compilation.

EasyCrypt~\cite{DBLP:conf/crypto/BartheGHB11} is a domain-specific
proof assistant that embeds many of the reasoning tools developed for
CertiCrypt. One key difference is that EasyCrypt is developed as a
standalone tool, rather than being embedded into an existing proof
assistant. EasyCrypt combines a proof engine for higher-order logic, a
backend to SMT\index{SMT solving} (satisfiability modulo theories\index{satisfiability modulo theories}) solvers, and support
for several program logics, including probabilistic Relational hoare
logic, and logics to upper-bound the probability of events,
including~\cite{DBLP:journals/pacml/AvanziniBGMV24}. Program logics
are ``natively'' embedded in the ambient logic: there is no
formalization of program semantics, and as a consequence one cannot
define the meaning of program logic judgments within the ambient
logic. This pragmatic approach eases experimenting with new logics;
for instance, EasyCrypt features a rich resource-aware module
system~\cite{DBLP:conf/ccs/BarbosaBGKS21}.  EasyCrypt has been used to
verify many examples from provable security, including encryption
schemes, signatures schemes \cite{DBLP:conf/csfw/FirsovLT21}, hash
functions, zero-knowledge protocols \cite{DBLP:conf/csfw/FirsovU23},
coin-tossing protocols \cite{DBLP:conf/cpp/FirsovU22}, distance
bounding protocols \cite{DBLP:conf/csfw/BoureanuDDG021}, e-voting
protocols \cite{DBLP:conf/sp/CortierDDSSW17}, multi-party protocols
\cite{DBLP:conf/csfw/StoughtonV17,DBLP:conf/csfw/HaaghKOSS18,DBLP:conf/csfw/SidorencoOS21,DBLP:conf/ccs/AlmeidaBCEG0P21},
and universal
composability~\cite{DBLP:conf/csfw/CanettiSV19,DBLP:conf/ccs/BarbosaBGKS21}. EasyCrypt
has also been used for verifying larger examples---e.g., a key
protocol of AWS Key Management
Service~\cite{DBLP:conf/ccs/AlmeidaBBCCGPPS19}.

\index{CertiCrypt|)}
\index{EasyCrypt|)}

\subsubsection{Foundational Cryptography Framework}

\index{Foundational Cryptography Framework|(}

Petcher and Morrisett~\cite{DBLP:conf/post/PetcherM15} use Rocq to formalize
the Foundational Cryptography Framework (FCF). In contrast to CertiCrypt, FCF
provides a shallow embedding of a probabilistic programming language, letting
users take advantage of the rich specification language of Rocq for writing
cryptographic constructions and security definitions. The two approaches
deliver different benefits in terms of expressiveness and automation and are
difficult to compare. FCF has been used to
mechanize a proof of security for a searchable symmetric encryption
scheme~\cite{DBLP:conf/csfw/PetcherM15} and for a proof of security of the HMAC
message authentication code~\cite{DBLP:conf/ccs/YeGSBPA17}.

\index{Foundational Cryptography Framework|)}

\subsubsection{CryptHOL}

\index{CryptHOL|(}

CryptHOL~\cite{DBLP:journals/joc/BasinLS20} is a formalization of
constructive cryptography\index{constructive cryptograph}~\cite{DBLP:conf/tosca/Maurer11}, a
foundational paradigm for compositional, simulation-based
security proofs in Isabelle\slash HOL.  
CryptHOL stands out from prior works such as CertiCrypt and
FCF, which are based on the code-based
game-based approach. CryptHOL's starting point is an encoding of
probabilistic interactive systems using coinductive types, instead of
the direct approach based on existential types. This encoding supports
all basic operators on probabilistic interactive systems, including
different forms of composition, and different notions of equivalence,
that can be used to reason about the security of cryptographic
protocols. CryptHOL also provides support for a relational program
logic akin to probabilistic relational Hoare logic, and for reasoning
principles such as optimistic sampling and up-to-bad reasoning, which
are widely used in security proofs. The framework is used to verify
indistinguishability of ElGamal, Hashed ElGamal, and other encryption
schemes.
Butler~\cite{DBLP:journals/jar/ButlerLAG21,DBLP:conf/cpp/BAG20,DBLP:conf/itp/Butler0G17}
use CryptHOL to verify $\Sigma$-protocols, commitment protocols,
oblivious transfer protocols, and secure two-party computations.

An extension~\cite{BLMS21} of the basic framework explores the interplay
between communication models and compositional security proofs. To this end,
the authors introduce the key notion of Fused Resource Template (FRT). At a
high-level, an FRT contains two parts:\ a core part and a rest part. The core
part describes the common behavior across different communication models, while
the rest part can interact with the core part in constrained ways. When
instantiating an FRT, one must ensure that the specification is respected; the
gain is that by doing so one obtains security guarantees automatically. This is
guaranteed by composition theorems for FRTs. The benefits of the approach are
illustrated through a running example on how to build a secure channel from a
Diffie--Hellmann key exchange protocol.

\index{CryptHOL|)}

\subsubsection{SSProve}

\index{SSProve|(}

SSProve~\cite{DBLP:journals/toplas/HaselwarterRMWASHMS23} is a Rocq library for
state-separating proofs\index{state-separating proofs}~\cite{DBLP:conf/asiacrypt/BrzuskaDFKK18}. A main idea
of state-separating proofs is to structure experiments using a notion of
package inspired from modules. The formalization provides proofs of the
algebraic laws of packages, and of the soundness of a relational program logic
for probabilistic computations. The framework is illustrated with ElGamal and
PRF-based encryption and key encapsulation mechanisms. A recent work by
Haselwarter {et al.}~\cite{DBLP:conf/cpp/HaselwarterHHWH24} establishes a
formal connection between SSProve and Jasmin\index{Jasmin}, and
use the resulting framework for proving security of PRF-based encryption.

\index{SSProve|)}

\subsubsection{Computational Indistinguishability Logic}

\index{Computational Indistinguishability Logic|(}

Corbineau, Duclos, and Lakhnech~\cite{CorbineauDL11} formalize Computational
Indistinguishability Logic~\cite{cil}; in contrast to prior works, this
formalization focuses on oracle systems and their interactions with
adversaries, without formalizing a programming language for describing
probabilistic computations. The formalization has been used to verify an
example of leakage-resilient cryptography.

\index{Computational Indistinguishability Logic|)}

\subsubsection{Interactive Probabilistic Dependency Logic}

\index{Interactive Probabilistic Dependency Logic|(}
\index{IPDL|(}

Gancher et al.~\cite{DBLP:journals/pacmpl/GancherSFSM23} introduce IPDL
(Interactive Probabilistic Dependency Logic), a Rocq
library that formalizes an equational logic\index{equational logic} to reason about distributed
probabilistic computations. In contrast to other formalizations, IPDL considers
communication channels explicitly. This treatment leads to a compositional
approach whereby properties of a protocol can be derived from its behavior
along communication channels. IPDL has been used to verify several examples,
including oblivious transfer and secure two-party computation.

\index{Interactive Probabilistic Dependency Logic|)}
\index{IPDL|)}

\subsubsection{Squirrel}

\index{Squirrel|(}

Squirrel~\cite{DBLP:conf/sp/BaeldeDJKM21} is a domain-specific proof assistant
tailored toward the Bana--Comon approach~\cite{banacomon}. The main idea of
this approach is to axiomatize the adversary's behavior in first-order logic.
However, rather than formalizing the adversary's capabilities, the approach is
based on specifying what the adversary cannot do (e.g., distinguish between two
ciphertexts). This leads to a notion of computationally complete symbolic
attacker. Squirrel has been used to verify a representative set of primitives.

\index{Squirrel|)}

\subsubsection{F$^{\star}$}

\index{F-star@$\text{F}^{\star}$|(}

Another alternative is to prove security of implementations using advanced
program verification tools. Such tools feature an intrinsic proof mode, based
on type-checking, and an extrinsic proof mode, which provides some basic
functionalities for interactive proofs. The main advantage of these tools is
that they integrate SMT solvers as backends, which can be used to discharge
proof obligations.

The Everest\index{Everest} project uses the $\text{F}^{\star}$
language~\cite{DBLP:conf/popl/SwamyHKRDFBFSKZ16} to build verified
implementations of cryptographic functions. $\text{F}^{\star}$ embeds a powerful refinement
type system, and some basic mechanisms for interactive proofs. This approach
carefully eschews probabilistic reasoning by replacing implementations of
primitives by deterministic functionalities. The approach primarily focuses on
trace properties, but support for relational reasoning is also
considered~\cite{DBLP:journals/pacmpl/MaillardHRM20}.

\index{F-star@$\text{F}^{\star}$|)}

\subsection{Security Proofs against Quantum Adversaries}

\index{post-quantum cryptography|(}
\index{quantum adversaries|(}

All of the formalizations and tools discussed so far consider a classic
execution model. However, there is an increasingly strong emphasis on
post-quantum cryptography---i.e., classical cryptography that resists quantum
adversaries and quantum cryptography. For instance, the National Institute of
Standards and Technology (NIST) is currently supervising a competition to
select and standardize a new set of cryptographic algorithms that can resist
quantum adversaries. Mechanizing security proofs of these algorithms involve
significant challenges, in particular adapting existing tools to the
(post-)quantum setting. One early work in this direction is
qRHL\index{qRHL}~\cite{Unruh19}, which is implemented in Isabelle/HOL. The formalization
has been used to prove security of the Fujasaki--Okamoto transform against
post-quantum adversaries~\cite{Unruh20}; this formalization is an important
step toward mechanizing security proofs of several NIST candidates. qRHL
supports reasoning about quantum programs. More recent projects focus on the
more specific goal of proving security of classical constructions against
quantum adversaries. This has the benefit of minimizing the gap with existing
tools; currently, EasyCrypt and Squirrel offer support for post-quantum
cryptography~\cite{DBLP:conf/ccs/BarbosaBFGHKSWZ21,CFJ-sp22}.

\index{post-quantum cryptography|)}
\index{quantum adversaries|)}

\subsection{Security Proofs in the Symbolic Model}

\index{symbolic model|(}

Groundbreaking work by Dolev and Yao~\cite{DBLP:journals/tit/DolevY83}
laid out the foundations for algorithmic verification of cryptographic
protocols. Their work defines a symbolic model of cryptography, where
cryptographic primitives\index{cryptography} (such as encryption and signatures) are
idealized and modeled purely algebraically. In this model, the
adversary can intercept, block, modify, or craft messages between
parties. These interactions give the adversary some knowledge that
they can exploit to recover cryptographic keys. Dolev and Yao show
that in their model security of a cryptographic protocol can be
decided in polynomial time. Following Lowe's
discovery~\cite{DBLP:journals/ipl/Lowe95} of a man-in-the-middle
attack on the Needham--Schr\"oder protocol, the Dolev--Yao\index{Dolev--Yao model} model has
been used extensively as a basis for formal verification of
cryptographic protocols~\cite{DBLP:conf/sp/BarbosaBBBCLP21}. Broadly
speaking, tools fall into two approaches: bounded tools, which
typically consider a finite number of sessions and perform state-space
exploration, and unbounded tools, which consider an infinite numbers
of sessions, and use a combination of approaches, including deductive
approaches. Early examples of unbounded tools include the NRL
analyzer~\cite{DBLP:journals/jlp/Meadows96},
ATHENA~\cite{DBLP:journals/jcs/SongBP01}. The NRL analyzer features
interactive and automated modes, whereas ATHENA is based on a custom
fully automated proof search procedure.

Influential early works by Bolignano~\cite{DBLP:conf/ccs/Bolignano96}
and Paulson~\cite{DBLP:journals/jcs/Paulson98} take the alternative
path to model the symbolic model in a proof assistant (Rocq and
Isabelle/HOL, respectively). The crux of their approach is to model
the knowledge of the adversary using an inductive relation. The
security of a protocol is then established by showing that at the end
of a protocol the knowledge of the adversary does not include secret
values. These approaches have been used to verify security properties
of real-world
protocols~\cite{DBLP:conf/csfw/Bolignano97,DBLP:conf/ccs/BellaPM02,DBLP:series/isc/Bella07}. Sprenger
et al.~\cite{DBLP:conf/csfw/SprengerBBPW06,DBLP:conf/csfw/SprengerB08}
formalize the Backes-Pfitzmann-Waidner model in Isabelle/HOL and use
their formalization to prove security of classic protocols, including
the (fixed) Needham-Schroeder protocol.
Goubault-Larrecq~\cite{DBLP:conf/csfw/Goubault-Larrecq08} studies the
problem of generating machine-checkable proofs from protocol
verification in the Dolev--Yao\index{Dolev--Yao model} model. Meier, Cremers, and
Basin~\cite{DBLP:conf/csfw/MeierCB10} implement proof-producing
procedure atop a shallow embedding of a protocol execution model in
Isabelle. Their procedure exploits protocol-independent invariants to
achieve automation.  Hess et al.~\cite{DBLP:conf/csfw/HessMBS21}
propose an automated approach for proving security of stateful
cryptographic protocols in Isabelle. Their approach is based on
abstract interpretation\index{abstract interpretation}, and computes a fixpoint that soundly
overapproximates protocol execution and can be checked automatically
for attacks. Braje et al.~\cite{DBLP:conf/csfw/BrajeLWKPKCC22}
formalize a model of cryptographic protocols with built-in safety
checks which ensure that trace properties can be verified without the
need to reason about attacker behavior.

Their model and the proof are formalized in Rocq. An alternative to
proving cryptographic protocols directly is to build secure protocols
by refinement. Sprenger and Basin~\cite{DBLP:conf/ccs/SprengerB10}
develop a refinement-based approach to reason about security of
cryptographic protocols in Isabelle. A series of follow-up works
instantiate their framework to obtain machine-checked security proofs
of key agreement under different
models~\cite{DBLP:conf/csfw/SprengerB12,DBLP:conf/eurosp/LallemandBS17}.
Klenze, Sprenger, and Basin~\cite{DBLP:conf/csfw/Klenze0B21} use a
similar approach to formalize the security of forwarding protocols in
Isabelle. Finally, Basin et al.~\cite{DBLP:journals/tissec/BasinCSS11}
and Cremers et al.~\cite{DBLP:conf/sp/CremersRSC12} develop extensions
of the basic model to reason about the security of physical and
distance bounding protocols in Isabelle.


\index{symbolic model|)}

\subsection{Security Proofs in the Generic Group Model}

\index{generic group model|(}

The generic group
model~\cite{DBLP:conf/eurocrypt/Shoup97,DBLP:conf/ima/Maurer05} is an idealized
model which can be used to reason about cryptographic constructions or problems
based on finite groups. One early application of the generic group model is to
prove generic lower bounds for solving the discrete logarithm problem. The
bounds hold for the restricted class of generic algorithms---i.e., algorithms
that do not have access to the group representation. The crux of the generic
group model is the Schwarz--Zippel lemma, which claims that the probability of
sampling uniformly at random a root of a multivariate polynomial $P$ of total
degree $d$ over a finite field $F$ is upper-bounded by $d/|F|$. Barthe,
Cederquist, and Tarento~\cite{DBLP:conf/cade/BartheCT04} formalize key results
of the generic group model in Rocq as well as several applications, including
lower bounds for solving the discrete logarithm, and proofs of ElGamal
encryption. 
model~\cite{DBLP:conf/crypto/FuchsbauerKL18} have found many novel
applications, including for zero-knowledge\index{zero-knowledge proofs} proofs.

\index{generic group model|)}

\subsection{Correctness Proofs}

There is a large body of work that formalizes mathematical concepts that arise
in cryptography. In particular, there exists several formalizations of elliptic\index{elliptic curves}
curves in Rocq \cite{DBLP:conf/tphol/TheryH07,DBLP:conf/itp/BartziaS14} and
Isabelle/HOL \cite{DBLP:conf/cade/HalesR20}. The latter formalizes Edwards
elliptic curves\index{elliptic curves}, which play a prominent role in recently proposed cryptographic
algorithms. There are also many works that use proof assistants for proving the
correctness of cryptographic implementations.

\subsubsection[$\mu$Cryptol]{$\pmb{\mu}$Cryptol}

Cryptol\index{Cryptol}, developed by Galois, 
is an embedded domain-specific language for
writing cryptographic algorithms. Cryptol offers support for checking
that programs are safe and that generated code is equivalent to its
Cryptol specification. Pike, Shields, and Matthews \cite{PikeSM06}
have also developed $\mu$Cryptol, a compiler from a fragment of the
Cryptol language to the AAMP7 microprocessor. The compiler is formally
verified in ACL2.

\subsubsection{Fiat-Crypto}
\label{ssec:fiat-crypto}

\index{Fiat-Crypto|(}

Erbsen et al.~\cite{DBLP:conf/sp/ErbsenPGSC19} use Rocq as a basis for
Fiat-Crypto, a compiler infrastructure to produce
correct-by-construction implementations of finite field and elliptic
curve cryptography. At a high level, Fiat-Crypto automatically
transforms mathematical descriptions of arithmetic\index{arithmetic} computations into
efficient, straight-line machine code\index{machine code}. Routines generated by FIAT
Crypto have been used as drop-in replacement of several previously
unverified routines in BoringSSL. Erbsen et al.~\cite{DBLP:journals/pacmpl/ErbsenPJLGPC24} use Fiat-Crypto
to verify a formally verified bare metal server that uses elliptic
curve cryptography. Hvass, Aranha, and
Spitters~\cite{DBLP:conf/csfw/HvassAS23} extend Fiat-Crypto to obtain
high-assurance implementations of field inversion.

In a different direction, Kuepper et
al.~\cite{DBLP:journals/pacmpl/KuepperEGCSTWCC23} combine Fiat-Crypto
with superoptimization techniques show further efficiency gains for
P-256 scalar multiplication. The correctness of their approach relies
on verified equivalence checking, which establishes semantic
equivalence between the algorithms output by Fiat-Crypto and the
algorithm output by the superoptimizer.

\index{Fiat-Crypto|)}

\subsubsection{Verified Software Toolchain}

\index{Verified Software Toolchain|(}

The Verified Software Toolchain (VST) projects uses a combination of the
CompCert\index{CompCert} verified compiler with a (formally verified) program
logic for C programs to verify cryptographic implementations. More
specifically, C implementations of SHA256 and HMAC are proved safe and correct
using a mechanization of separation logic\index{separation logic} built on top
of
CompCert\index{CompCert}~\cite{DBLP:journals/toplas/Appel15,DBLP:conf/uss/BeringerPYA15,
DBLP:conf/ccs/YeGSBPA17}; in addition to functional correctness, these
works establish reductionist security through a connection between
CompCert\index{CompCert} with FCF. More recently, Schwabe et al.\
use VST to establish the correctness of the TweetNaCl
implementation of Curve 25519~\cite{DBLP:conf/csfw/SchwabeVWW21}.

\index{Verified Software Toolchain|)}

\subsubsection{CryptoLine}

\index{CryptoLine|(}

CryptoLine is an automatic tool for verifying assembly implementations
of cryptographic routines. Early
work~\cite{DBLP:conf/ccs/ChenHLSTWYY14} uses an ad hoc combination of
SMT solvers and Rocq to verify the correctness of Curve 25519. This
approach is subsequently refined by Tsai, Wang, and
Yang~\cite{DBLP:conf/ccs/TsaiWY17}. Their refined approach transforms
verification tasks into modular polynomial equation entailment
problems, which can then be checked by computer algebra
systems. Solutions of the entailment problem are encoded into
certificates that are verified automatically in Rocq using computer-algebra-system-like
tactics. This line of work is further developed
in~\cite{DBLP:conf/cav/TsaiFLSWY23a}, where Tsai {et al.} present
COQCryptoLine\index{COQCryptoLine}, a variant of CryptoLine certified in Rocq. 

\index{CryptoLine|)}

\subsubsection{Jasmin}
\label{sssec:jasmin}

\index{Jasmin|(}

Jasmin~\cite{DBLP:conf/ccs/AlmeidaBBBGLOPS17,DBLP:conf/sp/AlmeidaBBGKL0S20}
is a framework that aims to deliver efficient, high-assurance
cryptographic implementations. The main components of the framework
are the Jasmin program verification infrastructure and the Jasmin
compiler. The Jasmin compiler is formally verified in Rocq, both for
safety and functional correctness.  This allows to reason about Jasmin
programs and obtain guarantees about assembly code. The Jasmin
verification infrastructure is based on EasyCrypt, via a translation
of Jasmin programs to EasyCrypt.

\index{Jasmin|)}

\subsubsection{HACL*}

\index{F-star@$\text{F}^{\star}$|(}
\index{HACL*|(}

The $\text{F}^{\star}$ language~\cite{DBLP:conf/popl/SwamyHKRDFBFSKZ16} and the $\text{F}^{\star}$/Vale
framework~\cite{DBLP:journals/pacmpl/FromherzGHPRS19} have been used to develop
the HACL* and EverCrypt libraries~\cite{ZinzindohoueBPB17,ProtzenkoPFHPBB20},
which have been widely deployed in popular systems.

\index{F-star@$\text{F}^{\star}$|)}
\index{HACL*|)}

\subsubsection{Other Approaches}

Ricketts el al.~\cite{DBLP:conf/pldi/RickettsRJTL14} formally verify a
SSH server in Rocq. Their formalization is based on Reflex\index{Reflex}, a deeply
embedded DSL with support for automated functional correctness and
noninterference proofs.

\index{cryptography|)}

\section{Other Applications}

\subsection{Zero-Knowledge and Electronic Voting}

\index{zero-knowledge proofs|(}
\index{electronic voting|(}
\index{voting|(}

Zero-knowledge proofs are cryptographic protocols that allow a prover $P$ to
prove knowledge of a secret to a verifier $V$, without revealing the secret to
$V$. The main properties of zero-knowledge proofs are soundness, completeness,
and zero-knowledge; the properties respectively (and informally) state that
cheating provers cannot convince verifiers, honest verifiers will accept
honestly generated proofs of valid statements, and verifiers will learn nothing
about the statement except its validity. Almeida et
al.~\cite{DBLP:conf/esorics/AlmeidaBBKSS10} use Isabelle/HOL as a backend to
generate soundness proofs for a subclass of zero-knowledge proofs known as
$\Sigma$-protocols. Subsequent work~\cite{DBLP:conf/ccs/AlmeidaBBBKB12} builds
a similar approach for a larger class of protocols, using a Rocq formalization
of proofs of knowledge of preimages under group homomorphisms~\cite{zkcc}. In
both cases, the formalizations are used as a backend by a cryptographic
compiler. The resulting certifying compilers take as input a logical formula
and generate a protocol together with a proof of its security properties:\
(special) soundness, completeness, and honest verifier zero-knowledge.

Haines and
collaborators~\cite{DBLP:conf/ccs/HainesGT19,DBLP:conf/sp/HainesGS21} use Rocq
to reason about correctness of verifiable mix nets used in electronic
elections. The proof involves reasoning about zero-knowledge arguments, which
is addressed through a careful modeling that eschews probabilistic reasoning.

\index{zero-knowledge proofs|)}
\index{electronic voting|)}
\index{voting|)}

\subsection{Smart Contracts}

\index{smart contracts|(}
\index{contracts|(}

Smart contracts are distributed programs that execute a protocol agreed by
several parties. Simple examples of smart contracts include swaps, lotteries,
payments, and other transactions. Smart contracts play an important role in
decentralized finance, and flaws in smart contracts can have devastating
financial consequences. It makes smart contracts, and the underlying
blockchains, an important target for formal verification.

Hirai~\cite{DBLP:conf/fc/Hirai17} formalizes the operational semantics
of the Ethereum VM in Rocq, HOL4, and Isabelle/HOL. Amani et al.~\cite{DBLP:conf/cpp/AmaniBBS18} formalize a
program logic for EVM bytecode\index{bytecode} in Isabelle/HOL and prove its soundness
with respect to Hirai's semantics. In a similar vein, Bernardo et
al.~\cite{DBLP:conf/fm/BernardoCHPT19} and Bernardo et
al.~\cite{DBLP:conf/isola/BernardoCCJPT20} formalize the operational
semantics and a sound program logic for Tezos smart contracts in
Rocq. There exists similar efforts to formalize intermediate or
low-level languages for smart contracts; for instance, there exists
formalized semantics of the Yul language in HOL4, Isabelle/HOL, and
Lean. More recently, Avigad et al.~\cite{DBLP:conf/cpp/AvigadGLST22}
propose an alternative approach to generate proofs of correctness for
algebraic programs written in the Cairo language.  Specifically, they
use Lean to show that Cairo programs whose
algebraic intermediate representations admit a solution have correct
executions.  Informally, these correct executions represent complete
runs of a protocol between a prover and a verifier. Their approach is
deployed to carry cryptocurrency exchanges.

In a different vein, Nielsen and Spitters~\cite{DBLP:conf/fm/NielsenS19} use
Rocq to reason about shallow embeddings of smart contracts. A more recent work
by Nielsen, Annenkov, and Spitters~\cite{DBLP:conf/cpp/NielsenAS23} use Rocq to
reason about decentralized exchanges. Applications to smart contracts have also
motivated mechanizations of consensus protocols. P\^irlea and
Sergey~\cite{DBLP:conf/cpp/PirleaS18} prove the correctness of consensus
protocols in Rocq. More generally, formalizations of distributed
protocols~\cite{DBLP:journals/pacmpl/SergeyWT18} could serve as a good starting
point to reason formally about smart contracts.

\index{smart contracts|)}
\index{contracts|)}

\subsection{Differential Privacy}

\index{differential privacy|(}

Differential privacy~\cite{DBLP:journals/fttcs/DworkR14} is a
quantitative, mathematically rigorous notion of privacy that
quantifies the amount of information leaked by a (randomized)
algorithm. Differential privacy is a relational property: informally,
an algorithm is differentially private if running the algorithm on two
closely related inputs yields closely related
distributions. Typically, inputs are databases, and two databases\index{databases} are
closely related (or, in differential privacy jargon, adjacent) if they
differ in one element. When elements are associated with individuals,
differential privacy thus guarantees that the information leaked about
a single individual is small. The magic of differential privacy is to
guarantee individual privacy while still allowing for the possibility
of statistically meaningful computations.

There are many notions of closeness for distributions; these notions yield
different notions of differential privacy, including vanilla differential
privacy (also known as $\epsilon$-differential privacy), approximate
differential privacy (also known as $\epsilon,\delta$-differential privacy),
and more recently R\'enyi differential privacy, which enjoys tighter
composition properties.

CertiPriv\index{CertiPriv}~\cite{DBLP:conf/popl/BartheKOB12} is a Rocq library to reason about
differential privacy. CertiPriv is built on top of CertiCrypt, and features an
approximate probabilistic relational Hoare logic, which is proved sound with
respect to a deep embedding of probabilistic programs. Later work uses
EasyCrypt in a similar style, for proving security of Sparse Vector, a
challenging algorithm from differential
privacy~\cite{DBLP:conf/ccs/BartheFGGHS16}.

\index{differential privacy|)}
\index{security|)}

{
\raggedright
\bibliographystyle{styles/spmpsci}
\bibliography{bib}

\begin{thebibliography}{100}
\providecommand{\url}[1]{{#1}}
\providecommand{\urlprefix}{URL }
\expandafter\ifx\csname urlstyle\endcsname\relax
  \providecommand{\doi}[1]{DOI~\discretionary{}{}{}#1}\else
  \providecommand{\doi}{DOI~\discretionary{}{}{}\begingroup
  \urlstyle{rm}\Url}\fi

\bibitem{cheri}
{CHERI} project.
\newblock \urlprefix\url{www.cheri-cpu.org}

\bibitem{DBLP:conf/icalp/Abadi98}
Abadi, M.: Protection in programming-language translations.
\newblock In: K.G. Larsen, S.~Skyum, G.~Winskel (eds.) ICALP '98, \emph{LNCS},
  vol. 1443, pp. 868--883. Springer (1998).
\newblock \urlprefix\url{https://doi.org/10.1007/BFb0055109}

\bibitem{DBLP:conf/ccs/AbateABEFHLPST18}
Abate, C., {Azevedo de Amorim}, A., Blanco, R., Evans, A.N., Fachini, G.,
  Hri\cb{t}cu, C., Laurent, T., Pierce, B.C., Stronati, M., Tolmach, A.: When
  good components go bad: Formally secure compilation despite dynamic
  compromise.
\newblock In: D.~Lie, M.~Mannan, M.~Backes, X.~Wang (eds.) CCS '18, pp.
  1351--1368. {ACM} (2018).
\newblock \urlprefix\url{https://doi.org/10.1145/3243734.3243745}

\bibitem{DBLP:journals/toplas/AbateBCDGHPTT21}
Abate, C., Blanco, R., Ciob{\^{a}}c{\u{a}}, {\c{S}}., Durier, A., Garg, D.,
  Hri\cb{t}cu, C., Patrignani, M., Tanter, {\'{E}}., Thibault, J.: An extended
  account of trace-relating compiler correctness and secure compilation.
\newblock {ACM} Trans. Program. Lang. Syst. \textbf{43}(4), 14:1--14:48 (2021).
\newblock \urlprefix\url{https://doi.org/10.1145/3460860}

\bibitem{DBLP:conf/csfw/AbateB0HPT19}
Abate, C., Blanco, R., Garg, D., Hri\cb{t}cu, C., Patrignani, M., Thibault, J.:
  Journey beyond full abstraction: Exploring robust property preservation for
  secure compilation.
\newblock In: CSF 2019, pp. 256--271. {IEEE} (2019).
\newblock \urlprefix\url{https://doi.org/10.1109/CSF.2019.00025}

\bibitem{DBLP:conf/provsec/AffeldtTM07}
Affeldt, R., Tanaka, M., Marti, N.: Formal proof of provable security by
  game-playing in a proof assistant.
\newblock In: W.~Susilo, J.K. Liu, Y.~Mu (eds.) ProvSec 2007, \emph{LNCS}, vol.
  4784, pp. 151--168. Springer (2007).
\newblock \urlprefix\url{https://doi.org/10.1007/978-3-540-75670-5\_10}

\bibitem{DBLP:conf/esorics/AlmeidaBBKSS10}
Almeida, J.B., Bangerter, E., Barbosa, M., Krenn, S., Sadeghi, A.R., Schneider,
  T.: A certifying compiler for zero-knowledge proofs of knowledge based on
  sigma-protocols.
\newblock In: D.~Gritzalis, B.~Preneel, M.~Theoharidou (eds.) ESORICS 2010,
  \emph{LNCS}, vol. 6345, pp. 151--167. Springer (2010).
\newblock \urlprefix\url{https://doi.org/10.1007/978-3-642-15497-3\_10}

\bibitem{DBLP:conf/ccs/AlmeidaBBBKB12}
Almeida, J.B., Barbosa, M., Bangerter, E., Barthe, G., Krenn, S.,
  B{\'{e}}guelin, S.Z.: Full proof cryptography: Verifiable compilation of
  efficient zero-knowledge protocols.
\newblock In: T.~Yu, G.~Danezis, V.D. Gligor (eds.) CCS '12, pp. 488--500.
  {ACM} (2012).
\newblock \urlprefix\url{https://doi.org/10.1145/2382196.2382249}

\bibitem{DBLP:conf/ccs/AlmeidaBBBGLOPS17}
Almeida, J.B., Barbosa, M., Barthe, G., Blot, A., Gr{\'{e}}goire, B., Laporte,
  V., Oliveira, T., Pacheco, H., Schmidt, B., Strub, P.Y.: Jasmin:
  High-assurance and high-speed cryptography.
\newblock In: B.M. Thuraisingham, D.~Evans, T.~Malkin, D.~Xu (eds.) {CCS} '17,
  pp. 1807--1823. {ACM} (2017).
\newblock \urlprefix\url{https://doi.org/10.1145/3133956.3134078}

\bibitem{DBLP:conf/ccs/AlmeidaBBCCGPPS19}
Almeida, J.B., Barbosa, M., Barthe, G., Campagna, M., Cohen, E.,
  Gr{\'{e}}goire, B., Pereira, V., Portela, B., Strub, P.Y., Tasiran, S.: A
  machine-checked proof of security for {AWS} key management service.
\newblock In: L.~Cavallaro, J.~Kinder, X.~Wang, J.~Katz (eds.) CCS '19, pp.
  63--78. {ACM} (2019).
\newblock \urlprefix\url{https://doi.org/10.1145/3319535.3354228}

\bibitem{DBLP:conf/ccs/AlmeidaBBD13}
Almeida, J.B., Barbosa, M., Barthe, G., Dupressoir, F.: Certified
  computer-aided cryptography: Efficient provably secure machine code from
  high-level implementations.
\newblock In: A.R. Sadeghi, V.D. Gligor, M.~Yung (eds.) CCS '13, pp.
  1217--1230. {ACM} (2013).
\newblock \urlprefix\url{https://doi.org/10.1145/2508859.2516652}

\bibitem{DBLP:conf/sp/AlmeidaBBGKL0S20}
Almeida, J.B., Barbosa, M., Barthe, G., Gr{\'{e}}goire, B., Koutsos, A.,
  Laporte, V., Oliveira, T., Strub, P.Y.: The last mile: High-assurance and
  high-speed cryptographic implementations.
\newblock In: SP 2020, pp. 965--982. {IEEE} (2020).
\newblock \urlprefix\url{https://doi.org/10.1109/SP40000.2020.00028}

\bibitem{DBLP:conf/ccs/AlmeidaBCEG0P21}
Almeida, J.B., Barbosa, M., Correia, M.L., Eldefrawy, K., Graham{-}Lengrand,
  S., Pacheco, H., Pereira, V.: Machine-checked {ZKP} for {NP} relations:
  Formally verified security proofs and implementations of mpc-in-the-head.
\newblock In: Y.~Kim, J.~Kim, G.~Vigna, E.~Shi (eds.) CCS '21, pp. 2587--2600.
  {ACM} (2021).
\newblock \urlprefix\url{https://doi.org/10.1145/3460120.3484771}

\bibitem{DBLP:conf/fopara/AmadioABBCGMMMPPRCST13}
Amadio, R.M., Ayache, N., Bobot, F., Boender, J., Campbell, B., Garnier, I.,
  Madet, A., McKinna, J., Mulligan, D.P., Piccolo, M., Pollack, R.,
  R{\'{e}}gis-Gianas, Y., Coen, C.S., Stark, I., Tranquilli, P.: {C}ertified
  {C}omplexity ({CerCo}).
\newblock In: U.D. Lago, R.~Pe{\~{n}}a (eds.) FOPARA 2013, \emph{LNCS}, vol.
  8552, pp. 1--18. Springer (2013).
\newblock \urlprefix\url{https://doi.org/10.1007/978-3-319-12466-7\_1}

\bibitem{DBLP:conf/cpp/AmaniBBS18}
Amani, S., B{\'{e}}gel, M., Bortin, M., Staples, M.: Towards verifying
  {Ethereum} smart contract bytecode in {Isabelle/HOL}.
\newblock In: J.~Andronick, A.P. Felty (eds.) CPP 2018, pp. 66--77. {ACM}
  (2018).
\newblock \urlprefix\url{https://doi.org/10.1145/3167084}

\bibitem{DBLP:conf/post/AmtoftDZABHOC12}
Amtoft, T., Dodds, J., Zhang, Z., Appel, A.W., Beringer, L., Hatcliff, J., Ou,
  X., Cousino, A.: A certificate infrastructure for machine-checked proofs of
  conditional information flow.
\newblock In: P.~Degano, J.D. Guttman (eds.) POST 2012, \emph{LNCS}, vol. 7215,
  pp. 369--389. Springer (2012).
\newblock \urlprefix\url{https://doi.org/10.1007/978-3-642-28641-4\_20}

\bibitem{DBLP:conf/tphol/AndronickCL03}
Andronick, J., Chetali, B., Ly, O.: Using {C}oq to verify {J}ava {C}ard applet
  isolation properties.
\newblock In: D.~Basin, B.~Wolff (eds.) TPHOLs 2003, \emph{LNCS}, vol. 2758,
  pp. 335--351. Springer (2003).
\newblock \urlprefix\url{https://doi.org/10.1007/10930755\_22}

\bibitem{DBLP:conf/fm/AndronickCP05}
Andronick, J., Chetali, B., Paulin{-}Mohring, C.: Formal verification of
  security properties of smart card embedded source code.
\newblock In: J.S. Fitzgerald, I.J. Hayes, A.~Tarlecki (eds.) FM 2005,
  \emph{LNCS}, vol. 3582, pp. 302--317. Springer (2005).
\newblock \urlprefix\url{https://doi.org/10.1007/11526841\_21}

\bibitem{DBLP:journals/pacmpl/AntonopoulosKLNNN23}
Antonopoulos, T., Koskinen, E., Le, T.C., Nagasamudram, R., Naumann, D.A., Ngo,
  M.: An algebra of alignment for relational verification.
\newblock Proc. {ACM} Program. Lang. \textbf{7}({POPL}), 573--603 (2023).
\newblock \urlprefix\url{https://doi.org/10.1145/3571213}

\bibitem{DBLP:journals/toplas/Appel15}
Appel, A.W.: Verification of a cryptographic primitive: {SHA-256}.
\newblock {ACM} Trans. Program. Lang. Syst. \textbf{37}(2), 7:1--7:31 (2015).
\newblock \urlprefix\url{https://doi.org/10.1145/2701415}

\bibitem{DBLP:conf/popl/AppelF00}
Appel, A.W., Felty, A.P.: A semantic model of types and machine instructions
  for proof-carrying code.
\newblock In: M.N. Wegman, T.W. Reps (eds.) {POPL} 2000, pp. 243--253. {ACM}
  (2000).
\newblock \urlprefix\url{https://doi.org/10.1145/325694.325727}

\bibitem{DBLP:journals/tcs/AspinallBHLM07}
Aspinall, D., Beringer, L., Hofmann, M., Loidl, H.W., Momigliano, A.: A program
  logic for resources.
\newblock Theor. Comput. Sci. \textbf{389}(3), 411--445 (2007).
\newblock \urlprefix\url{https://doi.org/10.1016/j.tcs.2007.09.003}

\bibitem{DBLP:journals/pacml/AvanziniBGMV24}
Avanzini, M., Barthe, G., Gr\'egoire, B., Moser, G., Vanoni, G.: Hopping proofs
  of expectation-based properties: Applications to skiplists and security
  proofs.
\newblock Proc. {ACM} Program. Lang. \textbf{8}({OOPSLA}) (2024)

\bibitem{DBLP:conf/cpp/AvigadGLST22}
Avigad, J., Goldberg, L., Levit, D., Seginer, Y., Titelman, A.: A verified
  algebraic representation of {C}airo program execution.
\newblock In: A.~Popescu, S.~Zdancewic (eds.) CPP '22, pp. 153--165. {ACM}
  (2022).
\newblock \urlprefix\url{https://doi.org/10.1145/3497775.3503675}

\bibitem{DBLP:journals/jcs/AmorimCDDHPPPT16}
{Azevedo de Amorim}, A., Collins, N., DeHon, A., Demange, D., Hri\cb{t}cu, C.,
  Pichardie, D., Pierce, B.C., Pollack, R., Tolmach, A.: A verified
  information-flow architecture.
\newblock J.~Comput. Secur. \textbf{24}(6), 689--734 (2016).
\newblock \urlprefix\url{https://doi.org/10.3233/JCS-15784}

\bibitem{DBLP:conf/sp/AmorimDGHPST15}
{Azevedo de Amorim}, A., D{\'{e}}n{\`{e}}s, M., Giannarakis, N., Hri\cb{t}cu,
  C., Pierce, B.C., Spector{-}Zabusky, A., Tolmach, A.: Micro-policies:
  Formally verified, tag-based security monitors.
\newblock In: SP 2015, pp. 813--830. {{IEEE}} (2015).
\newblock \urlprefix\url{https://doi.org/10.1109/SP.2015.55}

\bibitem{DBLP:conf/post/AmorimHP18}
{Azevedo de Amorim}, A., Hri\cb{t}cu, C., Pierce, B.C.: The meaning of memory
  safety.
\newblock In: L.~Bauer, R.~K{\"{u}}sters (eds.) POST 2018, \emph{LNCS}, vol.
  10804, pp. 79--105. Springer (2018).
\newblock \urlprefix\url{https://doi.org/10.1007/978-3-319-89722-6\_4}

\bibitem{DBLP:conf/sp/BaeldeDJKM21}
Baelde, D., Delaune, S., Jacomme, C., Koutsos, A., Moreau, S.: An interactive
  prover for protocol verification in the computational model.
\newblock In: {SP} 2021, pp. 537--554. {IEEE} (2021).
\newblock \urlprefix\url{https://doi.org/10.1109/SP40001.2021.00078}

\bibitem{banacomon}
Bana, G., Comon{-}Lundh, H.: A computationally complete symbolic attacker for
  equivalence properties.
\newblock In: G.J. Ahn, M.~Yung, N.~Li (eds.) CCS '14, pp. 609--620. {ACM}
  (2014).
\newblock \urlprefix\url{https://doi.org/10.1145/2660267.2660276}

\bibitem{DBLP:conf/sp/BarbosaBBBCLP21}
Barbosa, M., Barthe, G., Bhargavan, K., Blanchet, B., Cremers, C., Liao, K.,
  Parno, B.: {SoK}: Computer-aided cryptography.
\newblock In: {SP} 2021, pp. 777--795. {IEEE} (2021).
\newblock \urlprefix\url{https://doi.org/10.1109/SP40001.2021.00008}

\bibitem{DBLP:conf/ccs/BarbosaBFGHKSWZ21}
Barbosa, M., Barthe, G., Fan, X., Gr{\'{e}}goire, B., Hung, S.H., Katz, J.,
  Strub, P.Y., Wu, X., Zhou, L.: {EasyPQC}: Verifying post-quantum
  cryptography.
\newblock In: Y.~Kim, J.~Kim, G.~Vigna, E.~Shi (eds.) CCS '21, pp. 2564--2586.
  {ACM} (2021).
\newblock \urlprefix\url{https://doi.org/10.1145/3460120.3484567}

\bibitem{DBLP:conf/ccs/BarbosaBGKS21}
Barbosa, M., Barthe, G., Gr{\'{e}}goire, B., Koutsos, A., Strub, P.Y.:
  Mechanized proofs of adversarial complexity and application to universal
  composability.
\newblock In: Y.~Kim, J.~Kim, G.~Vigna, E.~Shi (eds.) CCS '21, pp. 2541--2563.
  {ACM} (2021).
\newblock \urlprefix\url{https://doi.org/10.1145/3460120.3484548}

\bibitem{DBLP:conf/fm/BartheBCL11}
Barthe, G., Betarte, G., Campo, J.D., Luna, C.: Formally verifying isolation
  and availability in an idealized model of virtualization.
\newblock In: M.J. Butler, W.~Schulte (eds.) FM 2011, \emph{LNCS}, vol. 6664,
  pp. 231--245. Springer (2011).
\newblock \urlprefix\url{https://doi.org/10.1007/978-3-642-21437-0\_19}

\bibitem{DBLP:conf/ccs/BartheBCLP14}
Barthe, G., Betarte, G., Campo, J.D., Luna, C.D., Pichardie, D.: System-level
  non-interference for constant-time cryptography.
\newblock In: G.J. Ahn, M.~Yung, N.~Li (eds.) CCS '14, pp. 1267--1279. {ACM}
  (2014).
\newblock \urlprefix\url{https://doi.org/10.1145/2660267.2660283}

\bibitem{DBLP:journals/pacmpl/BartheBGHLPT20}
Barthe, G., Blazy, S., Gr{\'{e}}goire, B., Hutin, R., Laporte, V., Pichardie,
  D., Trieu, A.: Formal verification of a constant-time preserving {C}
  compiler.
\newblock Proc. {ACM} Program. Lang. \textbf{4}({POPL}), 7:1--7:30 (2020).
\newblock \urlprefix\url{https://doi.org/10.1145/3371075}

\bibitem{DBLP:conf/cade/BartheCT04}
Barthe, G., Cederquist, J., Tarento, S.: A machine-checked formalization of the
  generic model and the random oracle model.
\newblock In: D.A. Basin, M.~Rusinowitch (eds.) IJCAR 2004, \emph{LNCS}, vol.
  3097, pp. 385--399. Springer (2004).
\newblock \urlprefix\url{https://doi.org/10.1007/978-3-540-25984-8\_29}

\bibitem{cil}
Barthe, G., Daubignard, M., Kapron, B.M., Lakhnech, Y.: Computational
  indistinguishability logic.
\newblock In: E.~Al{-}Shaer, A.D. Keromytis, V.~Shmatikov (eds.) CCS '10, pp.
  375--386. {ACM} (2010).
\newblock \urlprefix\url{https://doi.org/10.1145/1866307.1866350}

\bibitem{DBLP:conf/fase/BartheD04}
Barthe, G., Dufay, G.: A tool-assisted framework for certified bytecode
  verification.
\newblock In: M.~Wermelinger, T.~Margaria (eds.) FASE 2004, \emph{LNCS}, vol.
  2984, pp. 99--113. Springer (2004).
\newblock \urlprefix\url{https://doi.org/10.1007/978-3-540-24721-0\_7}

\bibitem{DBLP:conf/ccs/BartheFGGHS16}
Barthe, G., Fong, N., Gaboardi, M., Gr{\'{e}}goire, B., Hsu, J., Strub, P.Y.:
  Advanced probabilistic couplings for differential privacy.
\newblock In: E.R. Weippl, S.~Katzenbeisser, C.~Kruegel, A.C. Myers, S.~Halevi
  (eds.) CCS '16, pp. 55--67. {ACM} (2016).
\newblock \urlprefix\url{https://doi.org/10.1145/2976749.2978391}

\bibitem{DBLP:conf/popl/BartheGB09}
Barthe, G., Gr{\'{e}}goire, B., B{\'{e}}guelin, S.Z.: Formal certification of
  code-based cryptographic proofs.
\newblock In: Z.~Shao, B.C. Pierce (eds.) POPL 2009, pp. 90--101. {ACM} (2009).
\newblock \urlprefix\url{https://doi.org/10.1145/1480881.1480894}

\bibitem{DBLP:conf/crypto/BartheGHB11}
Barthe, G., Gr{\'{e}}goire, B., Heraud, S., B{\'{e}}guelin, S.Z.:
  Computer-aided security proofs for the working cryptographer.
\newblock In: P.~Rogaway (ed.) CRYPTO 2011, \emph{LNCS}, vol. 6841, pp. 71--90.
  Springer (2011).
\newblock \urlprefix\url{https://doi.org/10.1007/978-3-642-22792-9\_5}

\bibitem{DBLP:journals/jcs/BartheGHOB13}
Barthe, G., Gr{\'{e}}goire, B., Heraud, S., Olmedo, F., B{\'{e}}guelin, S.Z.:
  Verified indifferentiable hashing into elliptic curves.
\newblock J.~Comput. Secur. \textbf{21}(6), 881--917 (2013).
\newblock \urlprefix\url{https://doi.org/10.3233/JCS-130476}

\bibitem{DBLP:conf/csfw/BartheGL18}
Barthe, G., Gr{\'{e}}goire, B., Laporte, V.: Secure compilation of side-channel
  countermeasures: The case of cryptographic ``constant-time''.
\newblock In: CSF 2018, pp. 328--343. {{IEEE}} (2018).
\newblock \urlprefix\url{https://doi.org/10.1109/CSF.2018.00031}

\bibitem{DBLP:conf/ccs/BartheGLP21}
Barthe, G., Gr{\'{e}}goire, B., Laporte, V., Priya, S.: Structured leakage and
  applications to cryptographic constant-time and cost.
\newblock In: Y.~Kim, J.~Kim, G.~Vigna, E.~Shi (eds.) CCS '21, pp. 462--476.
  {ACM} (2021).
\newblock \urlprefix\url{https://doi.org/10.1145/3460120.3484761}

\bibitem{zkcc}
Barthe, G., Hedin, D., B{\'{e}}guelin, S.Z., Gr{\'{e}}goire, B., Heraud, S.: A
  machine-checked formalization of sigma-protocols.
\newblock In: CSF 2010, pp. 246--260. {{IEEE}} (2010).
\newblock \urlprefix\url{https://doi.org/10.1109/CSF.2010.24}

\bibitem{DBLP:conf/popl/BartheKOB12}
Barthe, G., K{\"{o}}pf, B., Olmedo, F., B{\'{e}}guelin, S.Z.: Probabilistic
  relational reasoning for differential privacy.
\newblock In: J.~Field, M.~Hicks (eds.) POPL 2012, pp. 97--110. {ACM} (2012).
\newblock \urlprefix\url{https://doi.org/10.1145/2103656.2103670}

\bibitem{DBLP:journals/jcs/BartheN07}
Barthe, G., Nieto, L.P.: Secure information flow for a concurrent language with
  scheduling.
\newblock J. Comput. Sec. \textbf{15}(6), 647--689 (2007).
\newblock
  \urlprefix\url{http://content.iospress.com/articles/journal-of-computer-security/jcs295}

\bibitem{DBLP:conf/esop/BarthePR07}
Barthe, G., Pichardie, D., Rezk, T.: A certified lightweight non-interference
  {J}ava bytecode verifier.
\newblock In: R.D. Nicola (ed.) ESOP 2007, \emph{LNCS}, vol. 4421, pp.
  125--140. Springer (2007).
\newblock \urlprefix\url{https://doi.org/10.1007/978-3-540-71316-6\_10}

\bibitem{DBLP:conf/itp/BartziaS14}
Bartzia, E.I., Strub, P.Y.: A formal library for elliptic curves in the {Coq}
  proof assistant.
\newblock In: G.~Klein, R.~Gamboa (eds.) ITP 2014, \emph{LNCS}, vol. 8558, pp.
  77--92. Springer (2014).
\newblock \urlprefix\url{https://doi.org/10.1007/978-3-319-08970-6\_6}

\bibitem{BLMS21}
Basin, D., Lochbihler, A., Maurer, U., Sefidgar, S.: Abstract modeling of
  system communication in constructive cryptography using {CryptHOL}.
\newblock In: CSF 2021, pp. 592--607. {{IEEE}} (2021).
\newblock \urlprefix\url{https://doi.org/10.1109/CSF51468.2021.00047}

\bibitem{DBLP:journals/tissec/BasinCSS11}
Basin, D.A., Capkun, S., Schaller, P., Schmidt, B.: Formal reasoning about
  physical properties of security protocols.
\newblock {ACM} Trans. Inf. Syst. Secur. \textbf{14}(2), 16:1--16:28 (2011).
\newblock \urlprefix\url{https://doi.org/10.1145/2019599.2019601}

\bibitem{DBLP:journals/joc/BasinLS20}
Basin, D.A., Lochbihler, A., Sefidgar, S.R.: {CryptHOL}: Game-based proofs in
  higher-order logic.
\newblock J.~Cryptol. \textbf{33}(2), 494--566 (2020).
\newblock \urlprefix\url{https://doi.org/10.1007/s00145-019-09341-z}

\bibitem{DBLP:conf/esop/BauereissCSAESB22}
Bauereiss, T., Campbell, B., Sewell, T., Armstrong, A., Esswood, L., Stark, I.,
  Barnes, G., Watson, R.N.M., Sewell, P.: Verified security for the morello
  capability-enhanced prototype arm architecture.
\newblock In: I.~Sergey (ed.) {ESOP} 2022, \emph{LNCS}, vol. 13240, pp.
  174--203. Springer (2022).
\newblock \urlprefix\url{https://doi.org/10.1007/978-3-030-99336-8\_7}

\bibitem{DBLP:conf/sp/BauereissG0R17}
Bauerei{\ss}, T., {Pesenti Gritti}, A., Popescu, A., Raimondi, F.: Cosmedis:
  {A} distributed social media platform with formally verified confidentiality
  guarantees.
\newblock In: SP 2017, pp. 729--748. {{IEEE}} (2017).
\newblock \urlprefix\url{https://doi.org/10.1109/SP.2017.24}

\bibitem{DBLP:series/isc/Bella07}
Bella, G.: Formal Correctness of Security Protocols.
\newblock \textit{Information Security and Cryptography}. Springer (2007).
\newblock \urlprefix\url{https://doi.org/10.1007/978-3-540-68136-6}

\bibitem{DBLP:conf/ccs/BellaPM02}
Bella, G., Paulson, L.C., Massacci, F.: The verification of an industrial
  payment protocol: The {SET} purchase phase.
\newblock In: V.~Atluri (ed.) CCS '02, pp. 12--20. {ACM} (2002).
\newblock \urlprefix\url{https://doi.org/10.1145/586110.586113}

\bibitem{DBLP:conf/eurocrypt/BellareR06}
Bellare, M., Rogaway, P.: The security of triple encryption and a framework for
  code-based game-playing proofs.
\newblock In: S.~Vaudenay (ed.) EUROCRYPT 2006, \emph{LNCS}, vol. 4004, pp.
  409--426. Springer (2006).
\newblock \urlprefix\url{https://doi.org/10.1007/11761679\_25}

\bibitem{DBLP:journals/jfp/Benzinger01}
Benzinger, R.: Automated complexity analysis of nuprl extracted programs
  journal of functional programming.
\newblock J.~Funct. Program. \textbf{11}(1), 3--31 (2001).
\newblock \urlprefix\url{https://doi.org/10.1017/s0956796800003865}

\bibitem{DBLP:conf/aplas/Beringer12}
Beringer, L.: End-to-end multilevel hybrid information flow control.
\newblock In: R.~Jhala, A.~Igarashi (eds.) APLAS 2012, \emph{LNCS}, vol. 7705,
  pp. 50--65. Springer (2012).
\newblock \urlprefix\url{https://doi.org/10.1007/978-3-642-35182-2\_5}

\bibitem{DBLP:conf/uss/BeringerPYA15}
Beringer, L., Petcher, A., Ye, K.Q., Appel, A.W.: Verified correctness and
  security of {OpenSSL} {HMAC}.
\newblock In: J.~Jung, T.~Holz (eds.) USENIX Security '15, pp. 207--221.
  {USENIX} Association (2015).
\newblock
  \urlprefix\url{https://www.usenix.org/conference/usenixsecurity15/technical-sessions/presentation/beringer}

\bibitem{DBLP:conf/isola/BernardoCCJPT20}
Bernardo, B., Cauderlier, R., Claret, G., Jakobsson, A., Pesin, B., Tesson, J.:
  Making {Tezos} smart contracts more reliable with {C}oq.
\newblock In: T.~Margaria, B.~Steffen (eds.) ISoLA 2020, Part III, \emph{LNCS},
  vol. 12478, pp. 60--72. Springer (2020).
\newblock \urlprefix\url{https://doi.org/10.1007/978-3-030-61467-6\_5}

\bibitem{DBLP:conf/fm/BernardoCHPT19}
Bernardo, B., Cauderlier, R., Hu, Z., Pesin, B., Tesson, J.: {Mi-Cho-Coq}, a
  framework for certifying {Tezos} smart contracts.
\newblock In: E.~Sekerinski, N.~Moreira, J.N. Oliveira, D.~Ratiu, R.~Guidotti,
  M.~Farrell, M.~Luckcuck, D.~Marmsoler, J.C. Campos, T.~Astarte, L.~Gonnord,
  A.~Cerone, L.~Couto, B.~Dongol, M.~Kutrib, P.~Monteiro, D.~Delmas (eds.) FM
  2019, Part I, \emph{LNCS}, vol. 12232, pp. 368--379. Springer (2019).
\newblock \urlprefix\url{https://doi.org/10.1007/978-3-030-54994-7\_28}

\bibitem{DBLP:conf/esop/BessonBDJW19}
Besson, F., Blazy, S., Dang, A., Jensen, T.P., Wilke, P.: Compiling sandboxes:
  Formally verified software fault isolation.
\newblock In: L.~Caires (ed.) ESOP 2019, \emph{LNCS}, vol. 11423, pp. 499--524.
  Springer (2019).
\newblock \urlprefix\url{https://doi.org/10.1007/978-3-030-17184-1\_18}

\bibitem{DBLP:conf/itp/BessonBW17}
Besson, F., Blazy, S., Wilke, P.: {CompCertS}: {A} memory-aware verified {C}
  compiler using pointer as integer semantics.
\newblock In: M.~Ayala{-}Rinc{\'{o}}n, C.A. Mu{\~{n}}oz (eds.) ITP 2017,
  \emph{LNCS}, vol. 10499, pp. 81--97. Springer (2017).
\newblock \urlprefix\url{https://doi.org/10.1007/978-3-319-66107-0\_6}

\bibitem{betarte2002formavie}
Betarte, G., Gim{\'e}nez, E., Loiseaux, C., Chetali, B.: {FORMAVIE}: Formal
  modelling and verification of the {JavaCard} 2.1.1 security architecture.
\newblock In: {e}-Smart 2002, pp. 213--231 (2002)

\bibitem{DBLP:conf/vstte/BlazyMP13}
Blazy, S., Maroneze, A.O., Pichardie, D.: Formal verification of loop bound
  estimation for {WCET} analysis.
\newblock In: E.~Cohen, A.~Rybalchenko (eds.) VSTTE 2013, \emph{LNCS}, vol.
  8164, pp. 281--303. Springer (2013).
\newblock \urlprefix\url{https://doi.org/10.1007/978-3-642-54108-7\_15}

\bibitem{Bohannon12}
Bohannon, A.: Foundations of web script security.
\newblock {PhD} thesis, University of Pennsylvania (2012)

\bibitem{DBLP:conf/ccs/BohannonPSWZ09}
Bohannon, A., Pierce, B.C., Sj{\"{o}}berg, V., Weirich, S., Zdancewic, S.:
  Reactive noninterference.
\newblock In: E.~Al{-}Shaer, S.~Jha, A.D. Keromytis (eds.) {CCS} '09, pp.
  79--90. {ACM} (2009).
\newblock \urlprefix\url{https://doi.org/10.1145/1653662.1653673}

\bibitem{DBLP:conf/ccs/Bolignano96}
Bolignano, D.: An approach to the formal verification of cryptographic
  protocols.
\newblock In: L.~Gong, J.~Stearn (eds.) CCS '96, pp. 106--118. {ACM} (1996).
\newblock \urlprefix\url{https://doi.org/10.1145/238168.238196}

\bibitem{DBLP:conf/csfw/Bolignano97}
Bolignano, D.: Towards the formal verification of electronic commerce
  protocols.
\newblock In: CSFW '97, pp. 133--147. {{IEEE}} (1997).
\newblock \urlprefix\url{https://doi.org/10.1109/CSFW.1997.596802}

\bibitem{DBLP:conf/csfw/BoureanuDDG021}
Boureanu, I., Dragan, C.C., Dupressoir, F., G{\'{e}}rault, D., Lafourcade, P.:
  Mechanised models and proofs for distance-bounding.
\newblock In: CSF 2021, pp. 1--16. {IEEE} (2021).
\newblock \urlprefix\url{https://doi.org/10.1109/CSF51468.2021.00049}

\bibitem{DBLP:journals/afp/BracevacGGMST18}
Bracevac, O., Gay, R., Grewe, S., Mantel, H., Sudbrock, H., Tasch, M.: An
  {I}sabelle/{HOL} formalization of the modular assembly kit for security
  properties.
\newblock Arch. Formal Proofs \textbf{2018} (2018).
\newblock
  \urlprefix\url{https://www.isa-afp.org/entries/Modular\_Assembly\_Kit\_Security.html}

\bibitem{DBLP:conf/csfw/BrajeLWKPKCC22}
Braje, T.M., Lee, A.R., Wagner, A., Kaiser, B., Park, D., Kalke, M.,
  Cunningham, R.K., Chlipala, A.: Adversary safety by construction in a
  language of cryptographic protocols.
\newblock In: CSF 2022, pp. 412--427. {IEEE} (2022).
\newblock \urlprefix\url{https://doi.org/10.1109/CSF54842.2022.9919638}

\bibitem{DBLP:journals/stvr/BruckerBW15}
Brucker, A.D., Br{\"{u}}gger, L., Wolff, B.: Formal firewall conformance
  testing: An application of test and proof techniques.
\newblock Softw. Test. Verif. Reliab. \textbf{25}(1), 34--71 (2015).
\newblock \urlprefix\url{https://doi.org/10.1002/stvr.1544}

\bibitem{DBLP:conf/asiacrypt/BrzuskaDFKK18}
Brzuska, C., Delignat{-}Lavaud, A., Fournet, C., Kohbrok, K., Kohlweiss, M.:
  State separation for code-based game-playing proofs.
\newblock In: T.~Peyrin, S.D. Galbraith (eds.) ASIACRYPT 2018, Part III,
  \emph{LNCS}, vol. 11274, pp. 222--249. Springer (2018).
\newblock \urlprefix\url{https://doi.org/10.1007/978-3-030-03332-3\_9}

\bibitem{DBLP:conf/itp/Butler0G17}
Butler, D., Aspinall, D., Gasc{\'{o}}n, A.: How to simulate it in {I}sabelle:
  Towards formal proof for secure multi-party computation.
\newblock In: M.~Ayala{-}Rinc{\'{o}}n, C.A. Mu{\~{n}}oz (eds.) ITP 2017,
  \emph{LNCS}, vol. 10499, pp. 114--130. Springer (2017).
\newblock \urlprefix\url{https://doi.org/10.1007/978-3-319-66107-0\_8}

\bibitem{DBLP:conf/cpp/BAG20}
Butler, D., Aspinall, D., Gasc{\'{o}}n, A.: Formalising oblivious transfer in
  the semi-honest and malicious model in {CryptHOL}.
\newblock In: J.~Blanchette, C.~Hri\cb{t}cu (eds.) CPP 2020, pp. 229--243.
  {ACM} (2020).
\newblock \urlprefix\url{https://doi.org/10.1145/3372885.3373815}

\bibitem{DBLP:journals/jar/ButlerLAG21}
Butler, D., Lochbihler, A., Aspinall, D., Gasc{\'{o}}n, A.: Formalising
  {$\Sigma$}-protocols and commitment schemes using {CryptHOL}.
\newblock J.~Autom. Reason. \textbf{65}(4), 521--567 (2021).
\newblock \urlprefix\url{https://doi.org/10.1007/s10817-020-09581-w}

\bibitem{DBLP:conf/fm/CacheraJPS05}
Cachera, D., Jensen, T.P., Pichardie, D., Schneider, G.: Certified memory usage
  analysis.
\newblock In: J.S. Fitzgerald, I.J. Hayes, A.~Tarlecki (eds.) FM 2005,
  \emph{LNCS}, vol. 3582, pp. 91--106. Springer (2005).
\newblock \urlprefix\url{https://doi.org/10.1007/11526841\_8}

\bibitem{DBLP:conf/csfw/CanettiSV19}
Canetti, R., Stoughton, A., Varia, M.: {EasyUC}: Using {EasyCrypt} to mechanize
  proofs of universally composable security.
\newblock In: CSF 2019, pp. 167--183. {IEEE} (2019).
\newblock \urlprefix\url{https://doi.org/10.1109/CSF.2019.00019}

\bibitem{DBLP:conf/ccs/CaprettaSFM07}
Capretta, V., Stepien, B., Felty, A.P., Matwin, S.: Formal correctness of
  conflict detection for firewalls.
\newblock In: P.~Ning, V.~Atluri, V.D. Gligor, H.~Mantel (eds.) FMSE 2007, pp.
  22--30. {ACM} (2007).
\newblock \urlprefix\url{https://doi.org/10.1145/1314436.1314440}

\bibitem{DBLP:conf/pldi/Carbonneaux0RS14}
Carbonneaux, Q., Hoffmann, J., Ramananandro, T., Shao, Z.: End-to-end
  verification of stack-space bounds for {C} programs.
\newblock In: M.F.P. O'Boyle, K.~Pingali (eds.) {PLDI} '14, pp. 270--281. {ACM}
  (2014).
\newblock \urlprefix\url{https://doi.org/10.1145/2594291.2594301}

\bibitem{DBLP:conf/cav/Carbonneaux0RS17}
Carbonneaux, Q., Hoffmann, J., Reps, T.W., Shao, Z.: Automated resource
  analysis with {C}oq proof objects.
\newblock In: R.~Majumdar, V.~Kuncak (eds.) CAV 2017, Part II, \emph{LNCS},
  vol. 10427, pp. 64--85. Springer (2017).
\newblock \urlprefix\url{https://doi.org/10.1007/978-3-319-63390-9\_4}

\bibitem{DBLP:conf/pldi/Carbonneaux0S15}
Carbonneaux, Q., Hoffmann, J., Shao, Z.: Compositional certified resource
  bounds.
\newblock In: D.~Grove, S.M. Blackburn (eds.) PLDI '15, pp. 467--478. {ACM}
  (2015).
\newblock \urlprefix\url{https://doi.org/10.1145/2737924.2737955}

\bibitem{DBLP:books/hal/Chargueraud23}
Chargu{\'{e}}raud, A.: A modern eye on separation logic for sequential
  programs.
\newblock Habilitation thesis (2023).
\newblock \urlprefix\url{https://tel.archives-ouvertes.fr/tel-04076725}

\bibitem{DBLP:conf/itp/ChargueraudP15}
Chargu{\'{e}}raud, A., Pottier, F.: Machine-checked verification of the
  correctness and amortized complexity of an efficient union-find
  implementation.
\newblock In: C.~Urban, X.~Zhang (eds.) ITP 2015, \emph{LNCS}, vol. 9236, pp.
  137--153. Springer (2015).
\newblock \urlprefix\url{https://doi.org/10.1007/978-3-319-22102-1\_9}

\bibitem{DBLP:journals/jar/ChargueraudP19}
Chargu{\'{e}}raud, A., Pottier, F.: Verifying the correctness and amortized
  complexity of a union-find implementation in separation logic with time
  credits.
\newblock J.~Autom. Reason. \textbf{62}(3), 331--365 (2019).
\newblock \urlprefix\url{https://doi.org/10.1007/s10817-017-9431-7}

\bibitem{DBLP:conf/ccs/ChenHLSTWYY14}
Chen, Y.F., Hsu, C.H., Lin, H.H., Schwabe, P., Tsai, M.H., Wang, B.Y., Yang,
  B.Y., Yang, S.Y.: Verifying {Curve25519} software.
\newblock In: G.J. Ahn, M.~Yung, N.~Li (eds.) CCS '14, pp. 299--309. {ACM}
  (2014).
\newblock \urlprefix\url{https://doi.org/10.1145/2660267.2660370}

\bibitem{DBLP:journals/jcs/ClarksonS10}
Clarkson, M.R., Schneider, F.B.: Hyperproperties.
\newblock J.~Comput. Secur. \textbf{18}(6), 1157--1210 (2010).
\newblock \urlprefix\url{https://doi.org/10.3233/JCS-2009-0393}

\bibitem{DBLP:journals/corr/abs-1211-6197}
Cock, D.A.: Verifying probabilistic correctness in isabelle with pgcl.
\newblock In: F.~Cassez, R.~Huuck, G.~Klein, B.~Schlich (eds.) SSV '12,
  \emph{{EPTCS}}, vol. 102, pp. 167--178 (2012).
\newblock \urlprefix\url{https://doi.org/10.4204/EPTCS.102.15}

\bibitem{CorbineauDL11}
Corbineau, P., Duclos, M., Lakhnech, Y.: Certified security proofs of
  cryptographic protocols in the computational model: An application to
  intrusion resilience.
\newblock In: J.P. Jouannaud, Z.~Shao (eds.) CPP 2011, \emph{LNCS}, vol. 7086,
  pp. 378--393. Springer (2011).
\newblock \urlprefix\url{https://doi.org/10.1007/978-3-642-25379-9_27}

\bibitem{DBLP:conf/sp/CortierDDSSW17}
Cortier, V., Dragan, C.C., Dupressoir, F., Schmidt, B., Strub, P.Y., Warinschi,
  B.: Machine-checked proofs of privacy for electronic voting protocols.
\newblock In: SP 2017, pp. 993--1008. {{IEEE}} (2017).
\newblock \urlprefix\url{https://doi.org/10.1109/SP.2017.28}

\bibitem{DBLP:conf/pldi/CostanzoSG16}
Costanzo, D., Shao, Z., Gu, R.: End-to-end verification of information-flow
  security for {C} and assembly programs.
\newblock In: C.~Krintz, E.D. Berger (eds.) PLDI '16, pp. 648--664. {ACM}
  (2016).
\newblock \urlprefix\url{https://doi.org/10.1145/2908080.2908100}

\bibitem{CFJ-sp22}
Cremers, C., Fontaine, C., Jacomme, C.: A logic and an interactive prover for
  the computational post-quantum security of protocols.
\newblock In: SP 2022, pp. 125--141. {IEEE} (2022).
\newblock \urlprefix\url{https://doi.org/10.1109/SP46214.2022.9833800}

\bibitem{DBLP:conf/sp/CremersRSC12}
Cremers, C.J.F., Rasmussen, K.B., Schmidt, B., Capkun, S.: Distance hijacking
  attacks on distance bounding protocols.
\newblock In: SP 2012, pp. 113--127. {{IEEE}} (2012).
\newblock \urlprefix\url{https://doi.org/10.1109/SP.2012.17}

\bibitem{DBLP:conf/sp/CriswellDA14}
Criswell, J., Dautenhahn, N., Adve, V.S.: Kcofi: Complete control-flow
  integrity for commodity operating system kernels.
\newblock In: SP 2014, pp. 292--307. {{IEEE}} (2014).
\newblock \urlprefix\url{https://doi.org/10.1109/SP.2014.26}

\bibitem{cutler2024cedar}
Cutler, J.W., Disselkoen, C., Eline, A., He, S., Headley, K., Hicks, M.,
  Hietala, K., Ioannidis, E., Kastner, J., Mamat, A., McAdams, D., McCutchen,
  M., Rungta, N., Torlak, E., Wells, A.: Cedar: A new language for expressive,
  fast, safe, and analyzable authorization (extended version).
\newblock Proc. {ACM} Program. Lang. \textbf{8}({OOPSLA}) (2024).
\newblock \urlprefix\url{https://doi.org/10.1145/3649835}

\bibitem{DBLP:conf/ccs/DamGKNS13}
Dam, M., Guanciale, R., Khakpour, N., Nemati, H., Schwarz, O.: Formal
  verification of information flow security for a simple arm-based separation
  kernel.
\newblock In: A.R. Sadeghi, V.D. Gligor, M.~Yung (eds.) {CCS} '13, pp.
  223--234. {ACM} (2013).
\newblock \urlprefix\url{https://doi.org/10.1145/2508859.2516702}

\bibitem{DBLP:conf/popl/Danielsson08}
Danielsson, N.A.: Lightweight semiformal time complexity analysis for purely
  functional data structures.
\newblock In: G.C. Necula, P.~Wadler (eds.) POPL 2008, pp. 133--144. {ACM}
  (2008).
\newblock \urlprefix\url{https://doi.org/10.1145/1328438.1328457}

\bibitem{DBLP:journals/pacml/pldi/DardinierM24}
Dardinier, T., M{\"{u}}ller, P.: {H}yper {H}oare {L}ogic: ({D}is-)proving
  program hyperproperties.
\newblock Proc. {ACM} Program. Lang. \textbf{8}({PLDI}) (2024).
\newblock \urlprefix\url{https://doi.org/10.48550/arXiv.2301.10037}

\bibitem{DBLP:journals/tit/DolevY83}
Dolev, D., Yao, A.C.C.: On the security of public key protocols.
\newblock IEEE Trans. Inf. Theory \textbf{29}(2), 198--207 (1983).
\newblock \urlprefix\url{https://doi.org/10.1109/TIT.1983.1056650}

\bibitem{DBLP:conf/sp/DSilvaPS15}
D'Silva, V., Payer, M., Song, D.X.: The correctness-security gap in compiler
  optimization.
\newblock In: 2015 {IEEE} Symposium on Security and Privacy Workshops, {SPW}
  2015, San Jose, CA, USA, May 21-22, 2015, pp. 73--87. {{IEEE}} (2015).
\newblock \urlprefix\url{https://doi.org/10.1109/SPW.2015.33}

\bibitem{DBLP:journals/fttcs/DworkR14}
Dwork, C., Roth, A.: The algorithmic foundations of differential privacy.
\newblock Found. Trends Theor. Comput. Sci. \textbf{9}(3--4), 211--407 (2014).
\newblock \urlprefix\url{https://doi.org/10.1561/0400000042}

\bibitem{DBLP:journals/jar/Eberl17}
Eberl, M.: Proving divide and conquer complexities in {I}sabelle/{HOL}.
\newblock J.~Autom. Reason. \textbf{58}(4), 483--508 (2017).
\newblock \urlprefix\url{https://doi.org/10.1007/s10817-016-9378-0}

\bibitem{DBLP:journals/jar/EberlHN20}
Eberl, M., Haslbeck, M.W., Nipkow, T.: Verified analysis of random binary tree
  structures.
\newblock J.~Autom. Reason. \textbf{64}(5), 879--910 (2020).
\newblock \urlprefix\url{https://doi.org/10.1007/s10817-020-09545-0}

\bibitem{DBLP:conf/csfw/El-KorashyBTDGH22}
El{-}Korashy, A., Blanco, R., Thibault, J., Durier, A., Garg, D., Hri\cb{t}cu,
  C.: {SecurePtrs}: Proving secure compilation with data-flow back-translation
  and turn-taking simulation.
\newblock In: CSF 2022, pp. 64--79. {IEEE} (2022).
\newblock \urlprefix\url{https://doi.org/10.1109/CSF54842.2022.9919680}

\bibitem{DBLP:conf/sp/ErbsenPGSC19}
Erbsen, A., Philipoom, J., Gross, J., Sloan, R., Chlipala, A.: Simple
  high-level code for cryptographic arithmetic---with proofs, without
  compromises.
\newblock In: {SP} 2019, pp. 1202--1219. {IEEE} (2019).
\newblock \urlprefix\url{https://doi.org/10.1109/SP.2019.00005}

\bibitem{DBLP:journals/pacmpl/ErbsenPJLGPC24}
Erbsen, A., Philipoom, J., Jamner, D., Lin, A., Gruetter, S., Pit{-}Claudel,
  C., Chlipala, A.: Foundational integration verification of a cryptographic
  server.
\newblock Proc. {ACM} Program. Lang. \textbf{8}({PLDI}), 1704--1729 (2024).
\newblock \urlprefix\url{https://doi.org/10.1145/3656446}

\bibitem{DBLP:conf/cav/ErnstM19}
Ernst, G., Murray, T.: {SecCSL}: Security concurrent separation logic.
\newblock In: I.~Dillig, S.~Tasiran (eds.) CAV 2019, \emph{LNCS}, vol. 11562,
  pp. 208--230. Springer (2019).
\newblock \urlprefix\url{https://doi.org/10.1007/978-3-030-25543-5\_13}

\bibitem{DBLP:journals/siglog/Finkbeiner23}
Finkbeiner, B.: Logics and algorithms for hyperproperties.
\newblock {ACM} {SIGLOG} News \textbf{10}(2), 4--23 (2023).
\newblock \urlprefix\url{https://doi.org/10.1145/3610392.3610394}

\bibitem{DBLP:conf/csfw/FirsovLT21}
Firsov, D., Lakk, H., Truu, A.: Verified multiple-time signature scheme from
  one-time signatures and timestamping.
\newblock In: CSF 2021, pp. 1--13. {IEEE} (2021).
\newblock \urlprefix\url{https://doi.org/10.1109/CSF51468.2021.00051}

\bibitem{DBLP:conf/cpp/FirsovU22}
Firsov, D., Unruh, D.: Reflection, rewinding, and coin-toss in {EasyCrypt}.
\newblock In: A.~Popescu, S.~Zdancewic (eds.) CPP '22, pp. 166--179. {ACM}
  (2022).
\newblock \urlprefix\url{https://doi.org/10.1145/3497775.3503693}

\bibitem{DBLP:conf/csfw/FirsovU23}
Firsov, D., Unruh, D.: Zero-knowledge in {EasyCrypt}.
\newblock In: CSF 2023, pp. 1--16. {IEEE} (2023).
\newblock \urlprefix\url{https://doi.org/10.1109/CSF57540.2023.00015}

\bibitem{DBLP:conf/csfw/FournetKL16}
Fournet, C., Keller, C., Laporte, V.: A certified compiler for verifiable
  computing.
\newblock In: {CSF} 2016, pp. 268--280. {{IEEE}} (2016).
\newblock \urlprefix\url{https://doi.org/10.1109/CSF.2016.26}

\bibitem{DBLP:journals/pacmpl/FromherzGHPRS19}
Fromherz, A., Giannarakis, N., Hawblitzel, C., Parno, B., Rastogi, A., Swamy,
  N.: A verified, efficient embedding of a verifiable assembly language.
\newblock Proc. {ACM} Program. Lang. \textbf{3}({POPL}), 63:1--63:30 (2019).
\newblock \urlprefix\url{https://doi.org/10.1145/3290376}

\bibitem{DBLP:conf/sp/FruminKB21}
Frumin, D., Krebbers, R., Birkedal, L.: Compositional non-interference for
  fine-grained concurrent programs.
\newblock In: SP 2021, pp. 1416--1433. {IEEE} (2021).
\newblock \urlprefix\url{https://doi.org/10.1109/SP40001.2021.00003}

\bibitem{DBLP:conf/crypto/FuchsbauerKL18}
Fuchsbauer, G., Kiltz, E., Loss, J.: The algebraic group model and its
  applications.
\newblock In: H.~Shacham, A.~Boldyreva (eds.) CRYPTO 2018, Part II,
  \emph{LNCS}, vol. 10992, pp. 33--62. Springer (2018).
\newblock \urlprefix\url{https://doi.org/10.1007/978-3-319-96881-0\_2}

\bibitem{DBLP:journals/pacmpl/GancherSFSM23}
Gancher, J., Sojakova, K., Fan, X., Shi, E., Morrisett, G.: A core calculus for
  equational proofs of cryptographic protocols.
\newblock Proc. {ACM} Program. Lang. \textbf{7}({POPL}), 866--892 (2023).
\newblock \urlprefix\url{https://doi.org/10.1145/3571223}

\bibitem{DBLP:journals/pacmpl/GeorgesGSTTHDB21}
Georges, A.L., Gu{\'{e}}neau, A., Strydonck, T.V., Timany, A., Trieu, A.,
  Huyghebaert, S., Devriese, D., Birkedal, L.: Efficient and provable local
  capability revocation using uninitialized capabilities.
\newblock Proc. {ACM} Program. Lang. \textbf{5}({POPL}), 1--30 (2021).
\newblock \urlprefix\url{https://doi.org/10.1145/3434287}

\bibitem{DBLP:journals/pacmpl/GeorgesTB22}
Georges, A.L., Trieu, A., Birkedal, L.: Le temps des cerises: Efficient
  temporal stack safety on capability machines using directed capabilities.
\newblock Proc. {ACM} Program. Lang. \textbf{6}({OOPSLA}), 1--30 (2022).
\newblock \urlprefix\url{https://doi.org/10.1145/3527318}

\bibitem{DBLP:journals/pacml/pldi/GladshteinZAAS24}
Gladshtein, V., Zhao, Q., Ahrens, W., Amarasinghe, S., Sergey, I.: Mechanised
  hypersafety proofs about structured data.
\newblock Proc. {ACM} Program. Lang. \textbf{8}({PLDI}) (2024).
\newblock \urlprefix\url{https://arxiv.org/abs/2404.06477}

\bibitem{DBLP:conf/sp/GoguenM82a}
Goguen, J.A., Meseguer, J.: Security policies and security models.
\newblock In: SP '82, pp. 11--20. {{IEEE}} (1982).
\newblock \urlprefix\url{https://doi.org/10.1109/SP.1982.10014}

\bibitem{DBLP:conf/sp/GoguenM84}
Goguen, J.A., Meseguer, J.: Unwinding and inference control.
\newblock In: SP '84, pp. 75--87. {{IEEE}} (1984).
\newblock \urlprefix\url{https://doi.org/10.1109/SP.1984.10019}

\bibitem{DBLP:journals/jcss/GoldwasserM84}
Goldwasser, S., Micali, S.: Probabilistic encryption.
\newblock J.~Comput. Syst. Sci. \textbf{28}(2), 270--299 (1984).
\newblock \urlprefix\url{https://doi.org/10.1016/0022-0000(84)90070-9}

\bibitem{DBLP:journals/pacmpl/Gomez-LondonoPS20}
G{\'{o}}mez{-}Londo{\~{n}}o, A., Pohjola, J.{\AA}., Syeda, H.T., Myreen, M.O.,
  Tan, Y.K.: Do you have space for dessert? {A} verified space cost semantics
  for {CakeML} programs.
\newblock Proc. {ACM} Program. Lang. \textbf{4}({OOPSLA}), 204:1--204:29
  (2020).
\newblock \urlprefix\url{https://doi.org/10.1145/3428272}

\bibitem{DBLP:journals/mscs/GorlaN16}
Gorla, D., Nestmann, U.: Full abstraction for expressiveness: History, myths
  and facts.
\newblock Math. Struct. Comput. Sci. \textbf{26}(4), 639--654 (2016).
\newblock \urlprefix\url{https://doi.org/10.1017/S0960129514000279}

\bibitem{DBLP:conf/csfw/Goubault-Larrecq08}
Goubault{-}Larrecq, J.: Towards producing formally checkable security proofs,
  automatically.
\newblock In: {CSF} 2008, pp. 224--238. {{IEEE}} (2008).
\newblock \urlprefix\url{https://doi.org/10.1109/CSF.2008.21}

\bibitem{DBLP:journals/pacmpl/GregersenBTB21}
Gregersen, S.O., Bay, J., Timany, A., Birkedal, L.: Mechanized logical
  relations for termination-insensitive noninterference.
\newblock Proc. {ACM} Program. Lang. \textbf{5}({POPL}), 1--29 (2021).
\newblock \urlprefix\url{https://doi.org/10.1145/3434291}

\bibitem{DBLP:conf/esop/GueneauCP18}
Gu{\'{e}}neau, A., Chargu{\'{e}}raud, A., Pottier, F.: A fistful of dollars:
  Formalizing asymptotic complexity claims via deductive program verification.
\newblock In: A.~Ahmed (ed.) ESOP 2018, \emph{LNCS}, vol. 10801, pp. 533--560.
  Springer (2018).
\newblock \urlprefix\url{https://doi.org/10.1007/978-3-319-89884-1\_19}

\bibitem{DBLP:conf/csfw/HaaghKOSS18}
Haagh, H., Karbyshev, A., Oechsner, S., Spitters, B., Strub, P.Y.:
  Computer-aided proofs for multiparty computation with active security.
\newblock In: CSF 2018, pp. 119--131. {{IEEE}} (2018).
\newblock \urlprefix\url{https://doi.org/10.1109/CSF.2018.00016}

\bibitem{DBLP:conf/sp/HainesGS21}
Haines, T., Gor{\'{e}}, R., Sharma, B.: Did you mix me? {F}ormally verifying
  verifiable mix nets in electronic voting.
\newblock In: {SP} 2021, pp. 1748--1765. {IEEE} (2021).
\newblock \urlprefix\url{https://doi.org/10.1109/SP40001.2021.00033}

\bibitem{DBLP:conf/ccs/HainesGT19}
Haines, T., Gor{\'{e}}, R., Tiwari, M.: Verified verifiers for verifying
  elections.
\newblock In: L.~Cavallaro, J.~Kinder, X.~Wang, J.~Katz (eds.) {CCS} '19, pp.
  685--702. {ACM} (2019).
\newblock \urlprefix\url{https://doi.org/10.1145/3319535.3354247}

\bibitem{DBLP:conf/cade/HalesR20}
Hales, T.C., Raya, R.: Formal proof of the group law for {E}dwards elliptic
  curves.
\newblock In: N.~Peltier, V.~Sofronie{-}Stokkermans (eds.) IJCAR 2020, Part II,
  \emph{LNCS}, vol. 12167, pp. 254--269. Springer (2020).
\newblock \urlprefix\url{https://doi.org/10.1007/978-3-030-51054-1\_15}

\bibitem{DBLP:journals/iacr/Halevi05a}
Halevi, S.: A plausible approach to computer-aided cryptographic proofs.
\newblock {IACR} Cryptol. ePrint Arch. \textbf{2005}(181) (2005).
\newblock \urlprefix\url{http://eprint.iacr.org/2005/181}

\bibitem{DBLP:conf/lics/HamidSTMN02}
Hamid, N.A., Shao, Z., Trifonov, V., Monnier, S., Ni, Z.: A syntactic approach
  to foundational proof-carrying code.
\newblock In: LICS 2002, pp. 89--100. {{IEEE}} (2002).
\newblock \urlprefix\url{https://doi.org/10.1109/LICS.2002.1029819}

\bibitem{DBLP:conf/acl2/HardinSY06}
Hardin, D.S., Smith, E.W., Young, W.D.: A robust machine code proof framework
  for highly secure applications.
\newblock In: P.~Manolios, M.~Wilding (eds.) ACL2 2006, pp. 11--20. {ACM}
  (2006).
\newblock \urlprefix\url{https://doi.org/10.1145/1217975.1217978}

\bibitem{DBLP:conf/cpp/HaselwarterHHWH24}
Haselwarter, P.G., Hvass, B.S., Hansen, L.L., Winterhalter, T., Hri\cb{t}cu,
  C., Spitters, B.: The last yard: Foundational end-to-end verification of
  high-speed cryptography.
\newblock In: A.~Timany, D.~Traytel, B.~Pientka, S.~Blazy (eds.) {CPP} 2024,
  pp. 30--44. {ACM} (2024).
\newblock \urlprefix\url{https://doi.org/10.1145/3636501.3636961}

\bibitem{DBLP:journals/toplas/HaselwarterRMWASHMS23}
Haselwarter, P.G., Rivas, E., Muylder, A.V., Winterhalter, T., Abate, C.,
  Sidorenco, N., Hri\cb{t}cu, C., Maillard, K., Spitters, B.: {SSProve}: {A}
  foundational framework for modular cryptographic proofs in {C}oq.
\newblock {ACM} Trans. Program. Lang. Syst. \textbf{45}(3), 15:1--15:61 (2023).
\newblock \urlprefix\url{https://doi.org/10.1145/3594735}

\bibitem{DBLP:conf/csfw/HessMBS21}
Hess, A.V., M{\"{o}}dersheim, S., Brucker, A.D., Schlichtkrull, A.: Performing
  security proofs of stateful protocols.
\newblock In: CSF 2021, pp. 1--16. {IEEE} (2021).
\newblock \urlprefix\url{https://doi.org/10.1109/CSF51468.2021.00006}

\bibitem{DBLP:conf/fc/Hirai17}
Hirai, Y.: Defining the {E}thereum virtual machine for interactive theorem
  provers.
\newblock In: M.~Brenner, K.~Rohloff, J.~Bonneau, A.~Miller, P.Y.A. Ryan,
  V.~Teague, A.~Bracciali, M.~Sala, F.~Pintore, M.~Jakobsson (eds.) FC 2017,
  \emph{LNCS}, vol. 10323, pp. 520--535. Springer (2017).
\newblock \urlprefix\url{https://doi.org/10.1007/978-3-319-70278-0\_33}

\bibitem{DBLP:conf/itp/Holzl16}
H{\"{o}}lzl, J.: Formalising semantics for expected running time of
  probabilistic programs.
\newblock In: J.C. Blanchette, S.~Merz (eds.) ITP 2016, \emph{LNCS}, vol. 9807,
  pp. 475--482. Springer (2016).
\newblock \urlprefix\url{https://doi.org/10.1007/978-3-319-43144-4\_30}

\bibitem{DBLP:journals/corr/abs-1212-3870}
H{\"{o}}lzl, J., Nipkow, T.: Interactive verification of {M}arkov chains: Two
  distributed protocol case studies.
\newblock In: U.~Fahrenberg, A.~Legay, C.R. Thrane (eds.) {QFM} 2012,
  \emph{{EPTCS}}, vol. 103, pp. 17--31 (2012).
\newblock \urlprefix\url{https://doi.org/10.4204/EPTCS.103.2}

\bibitem{DBLP:conf/sp/HritcuGKPM13}
Hri\cb{t}cu, C., Greenberg, M., Karel, B., Pierce, B.C., Morrisett, G.: All
  your {IFCException} are belong to us.
\newblock In: SP 2013, pp. 3--17. {{IEEE}} (2013).
\newblock \urlprefix\url{https://doi.org/10.1109/SP.2013.10}

\bibitem{DBLP:conf/csfw/HvassAS23}
Hvass, B.S., Aranha, D.F., Spitters, B.: High-assurance field inversion for
  curve-based cryptography.
\newblock In: CSF 2023, pp. 552--567. {IEEE} (2023).
\newblock \urlprefix\url{https://doi.org/10.1109/CSF57540.2023.00008}

\bibitem{DBLP:conf/uss/JangTL12}
Jang, D., Tatlock, Z., Lerner, S.: Establishing browser security guarantees
  through formal shim verification.
\newblock In: T.~Kohno (ed.) USENIX Security '12, pp. 113--128. {USENIX}
  Association (2012).
\newblock
  \urlprefix\url{https://www.usenix.org/conference/usenixsecurity12/technical-sessions/presentation/jang}

\bibitem{DBLP:journals/jacm/KaminskiKMO18}
Kaminski, B.L., Katoen, J.P., Matheja, C., Olmedo, F.: Weakest precondition
  reasoning for expected runtimes of randomized algorithms.
\newblock J.~{ACM} \textbf{65}(5), 30:1--30:68 (2018).
\newblock \urlprefix\url{https://doi.org/10.1145/3208102}

\bibitem{DBLP:conf/cav/KanavL014}
Kanav, S., Lammich, P., Popescu, A.: A conference management system with
  verified document confidentiality.
\newblock In: A.~Biere, R.~Bloem (eds.) CAV 2014, \emph{LNCS}, vol. 8559, pp.
  167--183. Springer (2014).
\newblock \urlprefix\url{https://doi.org/10.1007/978-3-319-08867-9\_11}

\bibitem{DBLP:journals/tcs/KleinN03}
Klein, G., Nipkow, T.: Verified bytecode verifiers.
\newblock Theor. Comput. Sci. \textbf{298}(3), 583--626 (2003).
\newblock \urlprefix\url{https://doi.org/10.1016/S0304-3975(02)00869-1}

\bibitem{DBLP:conf/csfw/Klenze0B21}
Klenze, T., Sprenger, C., Basin, D.A.: Formal verification of secure forwarding
  protocols.
\newblock In: CSF 2021, pp. 1--16. {IEEE} (2021).
\newblock \urlprefix\url{https://doi.org/10.1109/CSF51468.2021.00018}

\bibitem{DBLP:journals/pacmpl/KuepperEGCSTWCC23}
Kuepper, J., Erbsen, A., Gross, J., Conoly, O., Sun, C., Tian, S., Wu, D.,
  Chlipala, A., Chuengsatiansup, C., Genkin, D., Wagner, M., Yarom, Y.:
  {CryptOpt}: Verified compilation with randomized program search for
  cryptographic primitives.
\newblock Proc. {ACM} Program. Lang. \textbf{7}({PLDI}), 1268--1292 (2023).
\newblock \urlprefix\url{https://doi.org/10.1145/3591272}

\bibitem{DBLP:conf/eurosp/LallemandBS17}
Lallemand, J., Basin, D.A., Sprenger, C.: Refining authenticated key agreement
  with strong adversaries.
\newblock In: EuroS{\&}P 2017, pp. 92--107. {IEEE} (2017).
\newblock \urlprefix\url{https://doi.org/10.1109/EuroSP.2017.22}

\bibitem{DBLP:journals/cacm/Leroy09}
Leroy, X.: Formal verification of a realistic compiler.
\newblock Commun. {ACM} \textbf{52}(7), 107--115 (2009).
\newblock \urlprefix\url{https://doi.org/10.1145/1538788.1538814}

\bibitem{DBLP:conf/sp/LiLGNH21}
Li, S.W., Li, X., Gu, R., Nieh, J., Hui, J.Z.: A secure and formally verified
  {L}inux {KVM} hypervisor.
\newblock In: {SP} 2021, pp. 1782--1799. {IEEE} (2021).
\newblock \urlprefix\url{https://doi.org/10.1109/SP40001.2021.00049}

\bibitem{DBLP:journals/pacmpl/LiXW21}
Li, Y., yao Xia, L., Weirich, S.: Reasoning about the garden of forking paths.
\newblock Proc. {ACM} Program. Lang. \textbf{5}({ICFP}), 1--28 (2021).
\newblock \urlprefix\url{https://doi.org/10.1145/3473585}

\bibitem{DBLP:journals/ipl/Lowe95}
Lowe, G.: An attack on the {N}eedham--{S}chroeder public-key authentication
  protocol.
\newblock Inf. Process. Lett. \textbf{56}(3), 131--133 (1995).
\newblock \urlprefix\url{https://doi.org/10.1016/0020-0190(95)00144-2}

\bibitem{DBLP:journals/pacmpl/MaillardHRM20}
Maillard, K., Hri\cb{t}cu, C., Rivas, E., Muylder, A.V.: The next 700
  relational program logics.
\newblock Proc. {ACM} Program. Lang. \textbf{4}({POPL}), 4:1--4:33 (2020).
\newblock \urlprefix\url{https://doi.org/10.1145/3371072}

\bibitem{DBLP:phd/de/Mantel2004}
Mantel, H.: A uniform framework for the formal specification and verification
  of information flow security.
\newblock {PhD} thesis, Saarland University (2003).
\newblock
  \urlprefix\url{http://scidok.sulb.uni-saarland.de/volltexte/2004/202/index.html}

\bibitem{DBLP:conf/tosca/Maurer11}
Maurer, U.: Constructive cryptography---a new paradigm for security definitions
  and proofs.
\newblock In: S.~M{\"{o}}dersheim, C.~Palamidessi (eds.) {TOSCA} 2011,
  \emph{LNCS}, vol. 6993, pp. 33--56. Springer (2011).
\newblock \urlprefix\url{https://doi.org/10.1007/978-3-642-27375-9\_3}

\bibitem{DBLP:conf/ima/Maurer05}
Maurer, U.M.: Abstract models of computation in cryptography.
\newblock In: N.P. Smart (ed.) Cryptography and Coding 2005, \emph{LNCS}, vol.
  3796, pp. 1--12. Springer (2005).
\newblock \urlprefix\url{https://doi.org/10.1007/11586821\_1}

\bibitem{DBLP:journals/jlp/Meadows96}
Meadows, C.: The {NRL} {P}rotocol {A}nalyzer: An overview.
\newblock J.~Log. Program. \textbf{26}(2), 113--131 (1996).
\newblock \urlprefix\url{https://doi.org/10.1016/0743-1066(95)00095-X}

\bibitem{DBLP:conf/csfw/MeierCB10}
Meier, S., Cremers, C., Basin, D.A.: Strong invariants for the efficient
  construction of machine-checked protocol security proofs.
\newblock In: CSF 2010, pp. 231--245. {{IEEE}} (2010).
\newblock \urlprefix\url{https://doi.org/10.1109/CSF.2010.23}

\bibitem{DBLP:journals/scp/Mitchell93}
Mitchell, J.C.: On abstraction and the expressive power of programming
  languages.
\newblock Sci. Comput. Program. \textbf{21}(2), 141--163 (1993).
\newblock \urlprefix\url{https://doi.org/10.1016/0167-6423(93)90004-9}

\bibitem{DBLP:conf/cpp/Monniaux24}
Monniaux, D.: Memory simulations, security and optimization in a verified
  compiler.
\newblock In: A.~Timany, D.~Traytel, B.~Pientka, S.~Blazy (eds.) {CPP} 2024,
  pp. 103--117. {ACM} (2024).
\newblock \urlprefix\url{https://doi.org/10.1145/3636501.3636952}

\bibitem{DBLP:conf/pldi/MorrisettTTTG12}
Morrisett, G., Tan, G., Tassarotti, J., Tristan, J.B., Gan, E.: {RockSalt}:
  Better, faster, stronger {SFI} for the x86.
\newblock In: J.~Vitek, H.~Lin, F.~Tip (eds.) {PLDI} '12, pp. 395--404. {ACM}
  (2012).
\newblock \urlprefix\url{https://doi.org/10.1145/2254064.2254111}

\bibitem{DBLP:conf/sp/MurrayMBGBSLGK13}
Murray, T.C., Matichuk, D., Brassil, M., Gammie, P., Bourke, T., Seefried, S.,
  Lewis, C., Gao, X., Klein, G.: {seL4}: From general purpose to a proof of
  information flow enforcement.
\newblock In: SP 2013, pp. 415--429. {{IEEE}} (2013).
\newblock \urlprefix\url{https://doi.org/10.1109/SP.2013.35}

\bibitem{DBLP:conf/csfw/MurraySPR16}
Murray, T.C., Sison, R., Pierzchalski, E., Rizkallah, C.: Compositional
  verification and refinement of concurrent value-dependent noninterference.
\newblock In: CSF 2016, pp. 417--431. {{IEEE}} (2016).
\newblock \urlprefix\url{https://doi.org/10.1109/CSF.2016.36}

\bibitem{DBLP:conf/pldi/NagarakatteZMZ09}
Nagarakatte, S., Zhao, J., Martin, M.M.K., Zdancewic, S.: {SoftBound}: Highly
  compatible and complete spatial memory safety for {C}.
\newblock In: M.~Hind, A.~Diwan (eds.) PLDI '09, pp. 245--258. {ACM} (2009).
\newblock \urlprefix\url{https://doi.org/10.1145/1542476.1542504}

\bibitem{DBLP:conf/sp/NanevskiBG11}
Nanevski, A., Banerjee, A., Garg, D.: Verification of information flow and
  access control policies with dependent types.
\newblock In: SP 2011, pp. 165--179. {{IEEE}} (2011).
\newblock \urlprefix\url{https://doi.org/10.1109/SP.2011.12}

\bibitem{DBLP:conf/popl/Necula97}
Necula, G.C.: Proof-carrying code.
\newblock In: P.~Lee, F.~Henglein, N.D. Jones (eds.) POPL '97, pp. 106--119.
  {ACM} (1997).
\newblock \urlprefix\url{https://doi.org/10.1145/263699.263712}

\bibitem{DBLP:journals/sigops/NelsonBKTW20}
Nelson, L., Bornholt, J., Krishnamurthy, A., Torlak, E., Wang, X.:
  Noninterference specifications for secure systems.
\newblock {ACM} {SIGOPS} Oper. Syst. Rev. \textbf{54}(1), 31--39 (2020).
\newblock \urlprefix\url{https://doi.org/10.1145/3421473.3421478}

\bibitem{DBLP:conf/sp/NgoDFH17}
Ngo, V.C., Dehesa{-}Azuara, M., Fredrikson, M., Hoffmann, J.: Verifying and
  synthesizing constant-resource implementations with types.
\newblock In: SP 2017, pp. 710--728. {{IEEE}} (2017).
\newblock \urlprefix\url{https://doi.org/10.1109/SP.2017.53}

\bibitem{DBLP:conf/cpp/NielsenAS23}
Nielsen, E.H., Annenkov, D., Spitters, B.: Formalising decentralised exchanges
  in {C}oq.
\newblock In: R.~Krebbers, D.~Traytel, B.~Pientka, S.~Zdancewic (eds.) CPP
  2023, pp. 290--302. {ACM} (2023).
\newblock \urlprefix\url{https://doi.org/10.1145/3573105.3575685}

\bibitem{DBLP:conf/fm/NielsenS19}
Nielsen, J.B., Spitters, B.: Smart contract interactions in {C}oq.
\newblock In: E.~Sekerinski, N.~Moreira, J.N. Oliveira, D.~Ratiu, R.~Guidotti,
  M.~Farrell, M.~Luckcuck, D.~Marmsoler, J.C. Campos, T.~Astarte, L.~Gonnord,
  A.~Cerone, L.~Couto, B.~Dongol, M.~Kutrib, P.~Monteiro, D.~Delmas (eds.) FM
  2019, Part I, \emph{LNCS}, vol. 12232, pp. 380--391. Springer (2019).
\newblock \urlprefix\url{https://doi.org/10.1007/978-3-030-54994-7\_29}

\bibitem{DBLP:conf/sp/NienhuisJBFR0NN20}
Nienhuis, K., Joannou, A., Bauereiss, T., Fox, A.C.J., Roe, M., Campbell, B.,
  Naylor, M., Norton, R.M., Moore, S.W., Neumann, P.G., Stark, I., Watson,
  R.N.M., Sewell, P.: Rigorous engineering for hardware security: Formal
  modelling and proof in the {CHERI} design and implementation process.
\newblock In: SP 2020, pp. 1003--1020. {IEEE} (2020).
\newblock \urlprefix\url{https://doi.org/10.1109/SP40000.2020.00055}

\bibitem{DBLP:conf/itp/Nipkow15}
Nipkow, T.: Amortized complexity verified.
\newblock In: C.~Urban, X.~Zhang (eds.) ITP 2015, \emph{LNCS}, vol. 9236, pp.
  310--324. Springer (2015).
\newblock \urlprefix\url{https://doi.org/10.1007/978-3-319-22102-1\_21}

\bibitem{DBLP:conf/atva/NipkowEH20}
Nipkow, T., Eberl, M., Haslbeck, M.P.L.: Verified textbook algorithms: A biased
  survey.
\newblock In: D.V. Hung, O.~Sokolsky (eds.) ATVA 2020, \emph{LNCS}, vol. 12302,
  pp. 25--53. Springer (2020).
\newblock \urlprefix\url{https://doi.org/10.1007/978-3-030-59152-6\_2}

\bibitem{DBLP:conf/icics/Nowak07}
Nowak, D.: A framework for game-based security proofs.
\newblock In: S.~Qing, H.~Imai, G.~Wang (eds.) {ICICS} 2007, \emph{LNCS}, vol.
  4861, pp. 319--333. Springer (2007).
\newblock \urlprefix\url{https://doi.org/10.1007/978-3-540-77048-0\_25}

\bibitem{DBLP:conf/esorics/Oheimb04}
von Oheimb, D.: Information flow control revisited: Noninfluence $=$
  noninterference $+$ nonleakage.
\newblock In: P.~Samarati, P.Y.A. Ryan, D.~Gollmann, R.~Molva (eds.) {ESORICS}
  2004, \emph{LNCS}, vol. 3193, pp. 225--243. Springer (2004).
\newblock \urlprefix\url{https://doi.org/10.1007/978-3-540-30108-0\_14}

\bibitem{DBLP:journals/tches/OlmosBGGLLOS24}
Olmos, S.A., Barthe, G., Gonzalez, R., Gr{\'{e}}goire, B., Laporte, V.,
  L{\'{e}}chenet, J.C., Oliveira, T., Schwabe, P.: High-assurance zeroization.
\newblock {IACR} Trans. Cryptogr. Hardw. Embed. Syst. \textbf{2024}(1),
  375--397 (2024).
\newblock \urlprefix\url{https://doi.org/10.46586/tches.v2024.i1.375-397}

\bibitem{DBLP:journals/pacmpl/Paraskevopoulou19}
Paraskevopoulou, Z., Appel, A.W.: Closure conversion is safe for space.
\newblock Proc. {ACM} Program. Lang. \textbf{3}({ICFP}), 83:1--83:29 (2019).
\newblock \urlprefix\url{https://doi.org/10.1145/3341687}

\bibitem{DBLP:conf/tacas/ParkPM23}
Park, S.H., Pai, R.R., Melham, T.: A formal {CHERI-C} semantics for
  verification.
\newblock In: S.~Sankaranarayanan, N.~Sharygina (eds.) TACAS 2021, Part I,
  \emph{LNCS}, vol. 13993, pp. 549--568. Springer (2023).
\newblock \urlprefix\url{https://doi.org/10.1007/978-3-031-30823-9\_28}

\bibitem{DBLP:journals/mscs/Parrow16}
Parrow, J.: General conditions for full abstraction.
\newblock Math. Struct. Comput. Sci. \textbf{26}(4), 655--657 (2016).
\newblock \urlprefix\url{https://doi.org/10.1017/S0960129514000280}

\bibitem{DBLP:conf/csfw/PatrignaniG17}
Patrignani, M., Garg, D.: Secure compilation and hyperproperty preservation.
\newblock In: CSF 2017, pp. 392--404. {{IEEE}} (2017).
\newblock \urlprefix\url{https://doi.org/10.1109/CSF.2017.13}

\bibitem{DBLP:journals/jcs/Paulson98}
Paulson, L.C.: The inductive approach to verifying cryptographic protocols.
\newblock J. Comput. Sec. \textbf{6}(1--2), 85--128 (1998).
\newblock
  \urlprefix\url{http://content.iospress.com/articles/journal-of-computer-security/jcs102}

\bibitem{DBLP:conf/post/PetcherM15}
Petcher, A., Morrisett, G.: The foundational cryptography framework.
\newblock In: R.~Focardi, A.C. Myers (eds.) POST 2015, \emph{LNCS}, vol. 9036,
  pp. 53--72. Springer (2015).
\newblock \urlprefix\url{https://doi.org/10.1007/978-3-662-46666-7\_4}

\bibitem{DBLP:conf/csfw/PetcherM15}
Petcher, A., Morrisett, G.: A mechanized proof of security for searchable
  symmetric encryption.
\newblock In: C.~Fournet, M.W. Hicks, L.~Vigan{\`{o}} (eds.) CSF 2015, pp.
  481--494. {{IEEE}} (2015).
\newblock \urlprefix\url{https://doi.org/10.1109/CSF.2015.36}

\bibitem{PikeSM06}
Pike, L., Shields, M., Matthews, J.: A verifying core for a cryptographic
  language compiler.
\newblock In: P.~Manolios, M.~Wilding (eds.) ACL2 2006, pp. 1--10. {ACM}
  (2006).
\newblock \urlprefix\url{https://doi.org/10.1145/1217975.1217977}

\bibitem{DBLP:conf/cpp/PirleaS18}
P{\^{\i}}rlea, G., Sergey, I.: Mechanising blockchain consensus.
\newblock In: J.~Andronick, A.P. Felty (eds.) CPP 2018, pp. 78--90. {ACM}
  (2018).
\newblock \urlprefix\url{https://doi.org/10.1145/3167086}

\bibitem{DBLP:journals/tcs/Plotkin77}
Plotkin, G.D.: {LCF} considered as a programming language.
\newblock Theor. Comput. Sci. \textbf{5}(3), 223--255 (1977).
\newblock \urlprefix\url{https://doi.org/10.1016/0304-3975(77)90044-5}

\bibitem{DBLP:conf/cpp/PopescuHN12}
Popescu, A., H{\"{o}}lzl, J., Nipkow, T.: Proving concurrent noninterference.
\newblock In: C.~Hawblitzel, D.~Miller (eds.) CPP 2012, \emph{LNCS}, vol. 7679,
  pp. 109--125. Springer (2012).
\newblock \urlprefix\url{https://doi.org/10.1007/978-3-642-35308-6\_11}

\bibitem{DBLP:conf/cpp/0001HN13}
Popescu, A., H{\"{o}}lzl, J., Nipkow, T.: Formalizing probabilistic
  noninterference.
\newblock In: G.~Gonthier, M.~Norrish (eds.) CPP 2013, \emph{LNCS}, vol. 8307,
  pp. 259--275. Springer (2013).
\newblock \urlprefix\url{https://doi.org/10.1007/978-3-319-03545-1\_17}

\bibitem{DBLP:journals/pacmpl/PottierGJM24}
Pottier, F., Gu{\'{e}}neau, A., Jourdan, J.H., M{\'{e}}vel, G.: Thunks and
  debits in separation logic with time credits.
\newblock Proc. {ACM} Program. Lang. \textbf{8}({POPL}), 1482--1508 (2024).
\newblock \urlprefix\url{https://doi.org/10.1145/3632892}

\bibitem{ProtzenkoPFHPBB20}
Protzenko, J., Parno, B., Fromherz, A., Hawblitzel, C., Polubelova, M.,
  Bhargavan, K., Beurdouche, B., Choi, J., Delignat{-}Lavaud, A., Fournet, C.,
  Kulatova, N., Ramananandro, T., Rastogi, A., Swamy, N., Wintersteiger, C.M.,
  B{\'{e}}guelin, S.Z.: {EverCrypt}: {A} fast, verified, cross-platform
  cryptographic provider.
\newblock In: SP 2020, pp. 983--1002. {IEEE} (2020).
\newblock \urlprefix\url{https://doi.org/10.1109/SP40000.2020.00114}

\bibitem{DBLP:conf/pldi/RickettsRJTL14}
Ricketts, D., Robert, V., Jang, D., Tatlock, Z., Lerner, S.: Automating formal
  proofs for reactive systems.
\newblock In: M.F.P. O'Boyle, K.~Pingali (eds.) {PLDI} '14, pp. 452--462. {ACM}
  (2014).
\newblock \urlprefix\url{https://doi.org/10.1145/2594291.2594338}

\bibitem{Rushby92noninterference}
Rushby, J.: Noninterference, transitivity and channel-control security
  policies.
\newblock Tech. rep., SRI International (1992)

\bibitem{DBLP:journals/jsac/SabelfeldM03}
Sabelfeld, A., Myers, A.C.: Language-based information-flow security.
\newblock {IEEE} J.~Sel. Areas Commun. \textbf{21}(1), 5--19 (2003).
\newblock \urlprefix\url{https://doi.org/10.1109/JSAC.2002.806121}

\bibitem{DBLP:conf/csfw/SchwabeVWW21}
Schwabe, P., Viguier, B., Weerwag, T., Wiedijk, F.: A {Coq} proof of the
  correctness of {X25519} in {TweetNaCl}.
\newblock In: CSF 2021, pp. 1--16. {IEEE} (2021).
\newblock \urlprefix\url{https://doi.org/10.1109/CSF51468.2021.00023}

\bibitem{DBLP:journals/pacmpl/SergeyWT18}
Sergey, I., Wilcox, J.R., Tatlock, Z.: Programming and proving with distributed
  protocols.
\newblock Proc. {ACM} Program. Lang. \textbf{2}({POPL}), 28:1--28:30 (2018).
\newblock \urlprefix\url{https://doi.org/10.1145/3158116}

\bibitem{DBLP:conf/itp/SewellWGMAK11}
Sewell, T., Winwood, S., Gammie, P., Murray, T.C., Andronick, J., Klein, G.:
  {seL4} enforces integrity.
\newblock In: M.C.J.D. van Eekelen, H.~Geuvers, J.~Schmaltz, F.~Wiedijk (eds.)
  ITP 2011, \emph{LNCS}, vol. 6898, pp. 325--340. Springer (2011).
\newblock \urlprefix\url{https://doi.org/10.1007/978-3-642-22863-6\_24}

\bibitem{DBLP:conf/eurocrypt/Shoup97}
Shoup, V.: Lower bounds for discrete logarithms and related problems.
\newblock In: W.~Fumy (ed.) EUROCRYPT '97, \emph{LNCS}, vol. 1233, pp.
  256--266. Springer (1997).
\newblock \urlprefix\url{https://doi.org/10.1007/3-540-69053-0\_18}

\bibitem{DBLP:journals/iacr/Shoup04}
Shoup, V.: Sequences of games: A tool for taming complexity in security proofs.
\newblock {IACR} Cryptol. ePrint Arch. \textbf{2004}(332) (2004).
\newblock \urlprefix\url{http://eprint.iacr.org/2004/332}

\bibitem{DBLP:conf/csfw/SidorencoOS21}
Sidorenco, N., Oechsner, S., Spitters, B.: Formal security analysis of
  {MPC}-in-the-head zero-knowledge protocols.
\newblock In: CSF 2021, pp. 1--14. {IEEE} (2021).
\newblock \urlprefix\url{https://doi.org/10.1109/CSF51468.2021.00050}

\bibitem{DBLP:conf/ecoop/SilverHCHZ23}
Silver, L., He, P., Cecchetti, E., Hirsch, A.K., Zdancewic, S.: Semantics for
  noninterference with interaction trees.
\newblock In: K.~Ali, G.~Salvaneschi (eds.) ECOOP 2023, \emph{LIPIcs}, vol.
  263, pp. 29:1--29:29. Schloss Dagstuhl -- Leibniz-Zentrum f{\"{u}}r
  Informatik (2023).
\newblock \urlprefix\url{https://doi.org/10.4230/LIPIcs.ECOOP.2023.29}

\bibitem{DBLP:conf/eurosp/SimonCA18}
Simon, L., Chisnall, D., Anderson, R.J.: What you get is what you {C}:
  Controlling side effects in mainstream {C} compilers.
\newblock In: EuroS{\&}P 2018, pp. 1--15. {IEEE} (2018).
\newblock \urlprefix\url{https://doi.org/10.1109/EuroSP.2018.00009}

\bibitem{DBLP:journals/tkde/SohrDAG08}
Sohr, K., Drouineaud, M., Ahn, G.J., Gogolla, M.: Analyzing and managing
  role-based access control policies.
\newblock {IEEE} Trans. Knowl. Data Eng. \textbf{20}(7), 924--939 (2008).
\newblock \urlprefix\url{https://doi.org/10.1109/TKDE.2008.28}

\bibitem{DBLP:journals/jcs/SongBP01}
Song, D.X., Berezin, S., Perrig, A.: Athena: {A} novel approach to efficient
  automatic security protocol analysis.
\newblock J.~Comput. Secur. \textbf{9}(1/2), 47--74 (2001).
\newblock \urlprefix\url{https://doi.org/10.3233/jcs-2001-91-203}

\bibitem{DBLP:conf/csfw/SprengerBBPW06}
Sprenger, C., Backes, M., Basin, D.A., Pfitzmann, B., Waidner, M.:
  Cryptographically sound theorem proving.
\newblock In: CSFW '06, pp. 153--166. {{IEEE}} (2006).
\newblock \urlprefix\url{https://doi.org/10.1109/CSFW.2006.10}

\bibitem{DBLP:conf/csfw/SprengerB08}
Sprenger, C., Basin, D.A.: Cryptographically-sound protocol-model abstractions.
\newblock In: CSF 2008, pp. 115--129. {{IEEE}} (2008).
\newblock \urlprefix\url{https://doi.org/10.1109/CSF.2008.19}

\bibitem{DBLP:conf/ccs/SprengerB10}
Sprenger, C., Basin, D.A.: Developing security protocols by refinement.
\newblock In: E.~Al{-}Shaer, A.D. Keromytis, V.~Shmatikov (eds.) {CCS} '10, pp.
  361--374. {ACM} (2010).
\newblock \urlprefix\url{https://doi.org/10.1145/1866307.1866349}

\bibitem{DBLP:conf/csfw/SprengerB12}
Sprenger, C., Basin, D.A.: Refining key establishment.
\newblock In: S.~Chong (ed.) CSF 2012, pp. 230--246. {{IEEE}} (2012).
\newblock \urlprefix\url{https://doi.org/10.1109/CSF.2012.21}

\bibitem{DBLP:conf/cpp/St-MartinF16}
St{-}Martin, M., Felty, A.P.: A verified algorithm for detecting conflicts in
  {XACML} access control rules.
\newblock In: J.~Avigad, A.~Chlipala (eds.) CPP 2016, pp. 166--175. {ACM}
  (2016).
\newblock \urlprefix\url{https://doi.org/10.1145/2854065.2854079}

\bibitem{DBLP:conf/csfw/StoughtonV17}
Stoughton, A., Varia, M.: Mechanizing the proof of adaptive,
  information-theoretic security of cryptographic protocols in the random
  oracle model.
\newblock In: CSF 2017, pp. 83--99. {{IEEE}} (2017).
\newblock \urlprefix\url{https://doi.org/10.1109/CSF.2017.36}

\bibitem{DBLP:conf/csfw/StrydonckGGTTPB22}
Strydonck, T.V., Georges, A.L., Gu{\'{e}}neau, A., Trieu, A., Timany, A.,
  Piessens, F., Birkedal, L., Devriese, D.: Proving full-system security
  properties under multiple attacker models on capability machines.
\newblock In: CSF 2022, pp. 80--95. {IEEE} (2022).
\newblock \urlprefix\url{https://doi.org/10.1109/CSF54842.2022.9919645}

\bibitem{DBLP:conf/popl/SwamyHKRDFBFSKZ16}
Swamy, N., Hri\cb{t}cu, C., Keller, C., Rastogi, A., Delignat{-}Lavaud, A.,
  Forest, S., Bhargavan, K., Fournet, C., Strub, P.Y., Kohlweiss, M.,
  Zinzindohoue, J.K., B{\'{e}}guelin, S.Z.: Dependent types and multi-monadic
  effects in {F}.
\newblock In: R.~Bod{\'{\i}}k, R.~Majumdar (eds.) POPL 2016, pp. 256--270.
  {ACM} (2016).
\newblock \urlprefix\url{https://doi.org/10.1145/2837614.2837655}

\bibitem{DBLP:conf/sosp/TaoYLLNG21}
Tao, R., Yao, J., Li, X., Li, S.W., Nieh, J., Gu, R.: Formal verification of a
  multiprocessor hypervisor on {Arm} relaxed memory hardware.
\newblock In: R.~van Renesse, N.~Zeldovich (eds.) SOSP '21, pp. 866--881. {ACM}
  (2021).
\newblock \urlprefix\url{https://doi.org/10.1145/3477132.3483560}

\bibitem{DBLP:conf/itp/Tassarotti018}
Tassarotti, J., Harper, R.: Verified tail bounds for randomized programs.
\newblock In: J.~Avigad, A.~Mahboubi (eds.) ITP 2018, \emph{LNCS}, vol. 10895,
  pp. 560--578. Springer (2018).
\newblock \urlprefix\url{https://doi.org/10.1007/978-3-319-94821-8\_33}

\bibitem{DBLP:journals/pacmpl/TassarottiH19}
Tassarotti, J., Harper, R.: A separation logic for concurrent randomized
  programs.
\newblock Proc. {ACM} Program. Lang. \textbf{3}({POPL}), 64:1--64:30 (2019).
\newblock \urlprefix\url{https://doi.org/10.1145/3290377}

\bibitem{DBLP:conf/tphol/TheryH07}
Th{\'{e}}ry, L., Hanrot, G.: Primality proving with elliptic curves.
\newblock In: K.~Schneider, J.~Brandt (eds.) TPHOLs 2007, \emph{LNCS}, vol.
  4732, pp. 319--333. Springer (2007).
\newblock \urlprefix\url{https://doi.org/10.1007/978-3-540-74591-4\_24}

\bibitem{DBLP:conf/cav/TsaiFLSWY23a}
Tsai, M.H., Fu, Y.F., Liu, J., Shi, X., Wang, B.Y., Yang, B.Y.:
  {CoqCryptoLine}: {A} verified model checker with certified results.
\newblock In: C.~Enea, A.~Lal (eds.) CAV 2023, Part II, \emph{LNCS}, vol.
  13965, pp. 227--240. Springer (2023).
\newblock \urlprefix\url{https://doi.org/10.1007/978-3-031-37703-7\_11}

\bibitem{DBLP:conf/ccs/TsaiWY17}
Tsai, M.H., Wang, B.Y., Yang, B.Y.: Certified verification of algebraic
  properties on low-level mathematical constructs in cryptographic programs.
\newblock In: B.M. Thuraisingham, D.~Evans, T.~Malkin, D.~Xu (eds.) {CCS} '17,
  pp. 1973--1987. {ACM} (2017).
\newblock \urlprefix\url{https://doi.org/10.1145/3133956.3134076}

\bibitem{Unruh19}
Unruh, D.: Quantum relational {H}oare logic.
\newblock Proc. {ACM} Program. Lang. \textbf{3}({POPL}), 33:1--33:31 (2019).
\newblock \urlprefix\url{https://doi.org/10.1145/3290346}

\bibitem{Unruh20}
Unruh, D.: Post-quantum verification of {F}ujisaki-{O}kamoto.
\newblock In: S.~Moriai, H.~Wang (eds.) ASIACRYPT 2020, Part I, \emph{LNCS},
  vol. 12491, pp. 321--352. Springer (2020).
\newblock \urlprefix\url{https://doi.org/10.1007/978-3-030-64837-4\_11}

\bibitem{DBLP:journals/pacmpl/VassenaRGRS19}
Vassena, M., Russo, A., Garg, D., Rajani, V., Stefan, D.: From fine- to
  coarse-grained dynamic information flow control and back.
\newblock Proc. ACM Program. Lang. \textbf{3}({POPL}), 76:1--76:31 (2019).
\newblock \urlprefix\url{https://doi.org/10.1145/3290389}

\bibitem{DBLP:journals/pacmpl/WangWS19}
Wang, Y., Wilke, P., Shao, Z.: An abstract stack based approach to verified
  compositional compilation to machine code.
\newblock Proc. {ACM} Program. Lang. \textbf{3}({POPL}), 62:1--62:30 (2019).
\newblock \urlprefix\url{https://doi.org/10.1145/3290375}

\bibitem{DBLP:conf/pldi/WasserrabLS09}
Wasserrab, D., Lohner, D., Snelting, G.: On {PDG}-based noninterference and its
  modular proof.
\newblock In: S.~Chong, D.A. Naumann (eds.) {PLAS} 2009, pp. 31--44. {ACM}
  (2009).
\newblock \urlprefix\url{https://doi.org/10.1145/1554339.1554345}

\bibitem{DBLP:journals/pacmpl/XiaZHHMPZ20}
Xia, L., Zakowski, Y., He, P., Hur, C.K., Malecha, G., Pierce, B.C., Zdancewic,
  S.: Interaction trees: Representing recursive and impure programs in {C}oq.
\newblock Proc. {ACM} Program. Lang. \textbf{4}({POPL}), 51:1--51:32 (2020).
\newblock \urlprefix\url{https://doi.org/10.1145/3371119}

\bibitem{DBLP:conf/sp/XiangC21}
Xiang, J., Chong, S.: Co-inflow: Coarse-grained information flow control for
  {J}ava-like languages.
\newblock In: {SP} 2021, pp. 18--35. {IEEE} (2021).
\newblock \urlprefix\url{https://doi.org/10.1109/SP40001.2021.00002}

\bibitem{DBLP:conf/ccs/YeGSBPA17}
Ye, K.Q., Green, M., Sanguansin, N., Beringer, L., Petcher, A., Appel, A.W.:
  Verified correctness and security of {mbedTLS} {HMAC-DRBG}.
\newblock In: B.M. Thuraisingham, D.~Evans, T.~Malkin, D.~Xu (eds.) {CCS} '17,
  pp. 2007--2020. {ACM} (2017).
\newblock \urlprefix\url{https://doi.org/10.1145/3133956.3133974}

\bibitem{DBLP:conf/cav/YuanBTHZB22}
Yuan, S., Besson, F., Talpin, J.P., Hym, S., Zandberg, K., Baccelli, E.:
  End-to-end mechanized proof of an {eBPF} virtual machine for
  micro-controllers.
\newblock In: S.~Shoham, Y.~Vizel (eds.) CAV 2022, Part II, \emph{LNCS}, vol.
  13372, pp. 293--316. Springer (2022).
\newblock \urlprefix\url{https://doi.org/10.1007/978-3-031-13188-2\_15}

\bibitem{cheri-c-asplos}
Zaliva, V., Memarian, K., Almeida, R., Clarke, J., Davis, B., Richardson, A.,
  Chisnall, D., Campbell, B., Stark, I., Watson, R.N.M., Sewell, P.: Formal
  mechanised semantics of {CHERI} {C}: Capabilities, undefined behaviour, and
  provenance.
\newblock In: R.~Gupta, N.B. Abu{-}Ghazaleh, M.~Musuvathi, D.~Tsafrir (eds.)
  ASPLOS 2024, pp. 181--196. {ACM} (2024).
\newblock \urlprefix\url{https://doi.org/10.1145/3617232.3624859}

\bibitem{DBLP:conf/emsoft/ZhaoLSR11}
Zhao, L., Li, G., Sutter, B.D., Regehr, J.: {ARMor}: Fully verified software
  fault isolation.
\newblock In: S.~Chakraborty, A.~Jerraya, S.K. Baruah, S.~Fischmeister (eds.)
  {EMSOFT} 2011, pp. 289--298. {ACM} (2011).
\newblock \urlprefix\url{https://doi.org/10.1145/2038642.2038687}

\bibitem{ZinzindohoueBPB17}
Zinzindohou{\'{e}}, J.K., Bhargavan, K., Protzenko, J., Beurdouche, B.:
  {HACL*}: {A} verified modern cryptographic library.
\newblock In: B.M. Thuraisingham, D.~Evans, T.~Malkin, D.~Xu (eds.) CCS '17,
  pp. 1789--1806. {ACM} (2017).
\newblock \urlprefix\url{https://doi.org/10.1145/3133956.3134043}

\end{thebibliography}
}

\end{document}